\input harvmac.tex
\let\includefigures=\iftrue
\newfam\black
\includefigures
\input epsf
\def\figin{\epsfcheck\figin}\def\figins{\epsfcheck\figins}
\def\epsfcheck{\ifx\epsfbox\UnDeFiNeD
\message{(NO epsf.tex, FIGURES WILL BE IGNORED)}
\gdef\figin##1{\vskip2in}\gdef\figins##1{\hskip.5in}
\else\message{(FIGURES WILL BE INCLUDED)}%
\gdef\figin##1{##1}\gdef\figins##1{##1}\fi}
\def\DefWarn#1{}
\def\figinsert{\goodbreak\midinsert}
\def\ifig#1#2#3{\DefWarn#1\xdef#1{fig.~\the\figno}
\writedef{#1\leftbracket fig.\noexpand~\the\figno}%
\figinsert\figin{\centerline{#3}}\medskip\centerline{\vbox{\baselineskip12pt
\advance\hsize by -1truein\noindent\footnotefont{\bf Fig.~\the\figno:}
#2}}
\bigskip\endinsert\global\advance\figno by1}
\else
\def\ifig#1#2#3{\xdef#1{fig.~\the\figno}
\writedef{#1\leftbracket fig.\noexpand~\the\figno}%
#2}}
\global\advance\figno by1}
\fi


\def\sym{  \> {\vcenter  {\vbox
                  {\hrule height.6pt
                   \hbox {\vrule width.6pt  height5pt
                          \kern5pt
                          \vrule width.6pt  height5pt
                          \kern5pt
                          \vrule width.6pt height5pt}
                   \hrule height.6pt}
                             }
                  } \>
               }
\def\fund{  \> {\vcenter  {\vbox
                  {\hrule height.6pt
                   \hbox {\vrule width.6pt  height5pt
                          \kern5pt
                          \vrule width.6pt  height5pt }
                   \hrule height.6pt}
                             }
                       } \>
               }
\def\anti{ \>  {\vcenter  {\vbox
                  {\hrule height.6pt
                   \hbox {\vrule width.6pt  height5pt
                          \kern5pt
                          \vrule width.6pt  height5pt }
                   \hrule height.6pt
                   \hbox {\vrule width.6pt  height5pt
                          \kern5pt
                          \vrule width.6pt  height5pt }
                   \hrule height.6pt}
                             }
                  } \>
               }


\def\T{{\rm T}}

\def\la{\langle}
\def\ra{\rangle}
\def\al{\alpha}

\def\vac{{\rm vac}}

\def\T{{\rm T}}

\def\al{\alpha}
\def\si{\sigma}
\def\ga{\gamma}
\def\Ga{\Gamma}
\def\om{\omega}
\def\Om{\Omega}
\def\la{\langle}
\def\ra{\rangle}
\def\vac{{\rm vac}}
\def\w{\omega}
\def\bw{{\bar\omega}}


\lref\LeeCU{
P.~Lee and J.~w.~Park,
``Open strings in PP-wave background from defect conformal field theory,''
Phys.\ Rev.\ D {\bf 67}, 026002 (2003)
[arXiv:hep-th/0203257].
}
\lref\Bala{
V.~Balasubramanian, M.~x.~Huang, T.~S.~Levi and A.~Naqvi,
``Open strings from N = 4 super Yang-Mills,''
JHEP {\bf 0208}, 037 (2002)
[arXiv:hep-th/0204196].
}
\lref\Sken{
K.~Skenderis and M.~Taylor,
``Branes in AdS and pp-wave spacetimes,''
JHEP {\bf 0206}, 025 (2002)
[arXiv:hep-th/0204054].
}
\lref\SchwarzBC{
J.~H.~Schwarz,
``Comments on superstring interactions in a plane-wave background,''
JHEP {\bf 0209}, 058 (2002)
[arXiv:hep-th/0208179].
}

\lref\MannQP{
N.~Mann and J.~Polchinski,
``AdS Holography in the Penrose Limit,''
arXiv:hep-th/0305230.
}

\lref\ZhouMI{
J.~G.~Zhou,
``pp-wave string interactions from string bit model,''
arXiv:hep-th/0208232.
}

\lref\MetsaevBJ{
R.~R.~Metsaev,
``Type IIB Green-Schwarz superstring in plane wave Ramond-Ramond
background,''
Nucl.\ Phys.\ B {\bf 625}, 70 (2002)
[arXiv:hep-th/0112044].
}

\lref\GomisKJ{
J.~Gomis, S.~Moriyama and J.~w.~Park,
``SYM description of pp-wave string interactions: Singlet sector and
arbitrary impurities,''
arXiv:hep-th/0301250.
}

\lref\ConstableVQ{
N.~R.~Constable, D.~Z.~Freedman, M.~Headrick and S.~Minwalla,
``Operator mixing and the BMN correspondence,''
JHEP {\bf 0210}, 068 (2002)
[arXiv:hep-th/0209002].
}

\lref\LeeVZ{
P.~Lee, S.~Moriyama and J.~w.~Park,
``A note on cubic interactions in pp-wave light cone string field
theory,''
arXiv:hep-th/0209011.
}

\lref\largeN{
G.~'t Hooft,
``A Planar Diagram Theory For Strong Interactions,''
Nucl.\ Phys.\ B {\bf 72}, 461 (1974).
}

\lref\PankiewiczTG{
A.~Pankiewicz and B.~.~Stefanski,
``pp-wave light-cone superstring field theory,''
arXiv:hep-th/0210246.
}

\lref\PankiewiczGS{
A.~Pankiewicz,
``More comments on superstring interactions in the pp-wave background,''
JHEP {\bf 0209}, 056 (2002)
[arXiv:hep-th/0208209].
}

\lref\LeeRM{
P.~Lee, S.~Moriyama and J.~w.~Park,
``Cubic interactions in pp-wave light cone string field theory,''
Phys.\ Rev.\ D {\bf 66}, 085021 (2002)
[arXiv:hep-th/0206065].
}

\lref\MetsaevRE{
R.~R.~Metsaev and A.~A.~Tseytlin,
``Exactly solvable model of superstring in plane wave Ramond-Ramond
            background,''
Phys.\ Rev.\ D {\bf 65}, 126004 (2002)
[arXiv:hep-th/0202109].
}

\lref\DouglasJS{
M.~R.~Douglas, D.~A.~Lowe and J.~H.~Schwarz,
``Probing F-theory with multiple branes,''
Phys.\ Lett.\ B {\bf 394}, 297 (1997)
[arXiv:hep-th/9612062].
}

\lref\BlauNE{
M.~Blau, J.~Figueroa-O'Farrill, C.~Hull and G.~Papadopoulos,
``A new maximally supersymmetric background of IIB superstring
theory,''
JHEP {\bf 0201}, 047 (2002)
[arXiv:hep-th/0110242].
}

\lref\BerensteinJQ{
D.~Berenstein, J.~M.~Maldacena and H.~Nastase,
``Strings in flat space and pp waves from N = 4 super Yang Mills,''
JHEP {\bf 0204}, 013 (2002)
[arXiv:hep-th/0202021].
}

\lref\BerensteinZW{
D.~Berenstein, E.~Gava, J.~M.~Maldacena, K.~S.~Narain and H.~Nastase,
``Open strings on plane waves and their Yang-Mills duals,''
arXiv:hep-th/0203249.
}

\lref\SpradlinAR{
M.~Spradlin and A.~Volovich,
``Superstring interactions in a pp-wave background,''
Phys.\ Rev.\ D {\bf 66}, 086004 (2002)
[arXiv:hep-th/0204146].
}

\lref\SpradlinRV{
M.~Spradlin and A.~Volovich,
``Superstring interactions in a pp-wave background. II,''
arXiv:hep-th/0206073.
}

\lref\RoibanXR{
R.~Roiban, M.~Spradlin and A.~Volovich,
``On light-cone SFT contact terms in a plane wave,''
arXiv:hep-th/0211220.
}

\lref\AharonyXZ{
O.~Aharony, A.~Fayyazuddin and J.~M.~Maldacena,
``The large N limit of N = 2,1 field theories from three-branes in
F-theory,''
JHEP {\bf 9807}, 013 (1998)
[arXiv:hep-th/9806159].
}

\lref\GrossSU{
D.~J.~Gross, A.~Mikhailov and R.~Roiban,
``Operators with large R charge in N = 4 Yang-Mills theory,''
Annals Phys.\  {\bf 301}, 31 (2002)
[arXiv:hep-th/0205066].
}

\lref\SantambrogioSB{
A.~Santambrogio and D.~Zanon,
``Exact anomalous dimensions of N = 4 Yang-Mills operators with large
R  charge,''
Phys.\ Lett.\ B {\bf 545}, 425 (2002)
[arXiv:hep-th/0206079].
}

\lref\GrossMH{
D.~J.~Gross, A.~Mikhailov and R.~Roiban,
``A calculation of the plane wave string Hamiltonian from N = 4
super-Yang-Mills theory,''
arXiv:hep-th/0208231.
}

\lref\GomisWI{
J.~Gomis, S.~Moriyama and J.~w.~Park,
``SYM description of SFT Hamiltonian in a pp-wave background,''
arXiv:hep-th/0210153.
}

\lref\BerensteinSA{
D.~Berenstein and H.~Nastase,
``On lightcone string field theory from super Yang-Mills and
holography,''
arXiv:hep-th/0205048.
}

\lref\PearsonZS{
J.~Pearson, M.~Spradlin, D.~Vaman, H.~Verlinde and A.~Volovich,
``Tracing the string: BMN correspondence at finite J**2/N,''
arXiv:hep-th/0210102.
}

\lref\ImamuraWZ{
Y.~Imamura,
``Open string: BMN operator correspondence in the weak coupling regime,''
Prog.\ Theor.\ Phys.\  {\bf 108}, 1077 (2003)
[arXiv:hep-th/0208079].
}

\lref\BlauDY{
M.~Blau, J.~Figueroa-O'Farrill, C.~Hull and G.~Papadopoulos,
``Penrose limits and maximal supersymmetry,''
Class.\ Quant.\ Grav.\  {\bf 19}, L87 (2002)
[arXiv:hep-th/0201081].
}

\lref\VamanKA{
D.~Vaman and H.~Verlinde,
``Bit strings from N = 4 gauge theory,''
arXiv:hep-th/0209215.
}

\lref\VerlindeIG{
H.~Verlinde,
``Bits, matrices and 1/N,''
arXiv:hep-th/0206059.
}

\lref\ChandrasekharFQ{
B.~Chandrasekhar and A.~Kumar,
``D-branes in pp-wave light cone string field theory,''
arXiv:hep-th/0303223.
}

\lref\ParnachevKK{
A.~Parnachev and A.~V.~Ryzhov,
``Strings in the near plane wave background and AdS/CFT,''
JHEP {\bf 0210}, 066 (2002)
[arXiv:hep-th/0208010].
}

\lref\HokerTZ{
E.~D'Hoker, D.~Z.~Freedman and W.~Skiba,
``Field theory tests for correlators in the AdS/CFT correspondence,''
Phys.\ Rev.\ D {\bf 59}, 045008 (1999)
[arXiv:hep-th/9807098].
}

\lref\BianchiRW{
M.~Bianchi, B.~Eden, G.~Rossi and Y.~S.~Stanev,
``On operator mixing in N = 4 SYM,''
Nucl.\ Phys.\ B {\bf 646}, 69 (2002)
[arXiv:hep-th/0205321].
}

\lref\GreenTC{
M.~B.~Green and J.~H.~Schwarz,
``Superstring Interactions,''
Nucl.\ Phys.\ B {\bf 218}, 43 (1983).
}

\lref\GreenQU{
M.~B.~Green and J.~H.~Schwarz,
``The Structure Of Superstring Field Theories,''
Phys.\ Lett.\ B {\bf 140}, 33 (1984).
}

\lref\GreenFU{
M.~B.~Green and J.~H.~Schwarz,
``Superstring Field Theory,''
Nucl.\ Phys.\ B {\bf 243}, 475 (1984).}

\lref\GreenHW{
M.~B.~Green, J.~H.~Schwarz and L.~Brink,
``Superfield Theory Of Type Ii Superstrings,''
Nucl.\ Phys.\ B {\bf 219}, 437 (1983).}

\lref\NaculichFH{
S.~G.~Naculich, H.~J.~Schnitzer and N.~Wyllard,
``pp-wave limits and orientifolds,''
Nucl.\ Phys.\ B {\bf 650}, 43 (2003)
[arXiv:hep-th/0206094].
}

\lref\GursoyFJ{
U.~Gursoy,
``Predictions for pp-wave string amplitudes from perturbative SYM,''
arXiv:hep-th/0212118.
}

\lref\StefanskiZC {
B.~.~Stefanski,
``Open string plane-wave light-cone superstring field theory,''
arXiv:hep-th/0304114.
}

\Title{\vbox{\baselineskip12pt\hbox{CALT-68-2442}\hbox{hep-th/0305264}}}
{\vbox{\centerline{Open+Closed String Field Theory From Gauge Fields}}}

\centerline{Jaume Gomis, Sanefumi Moriyama and Jongwon
Park\footnote{$^\dagger$}
{gomis, moriyama, jongwon@theory.caltech.edu}}
\bigskip\centerline{\it California Institute of Technology 452-48,
Pasadena, CA 91125}
\vskip .3in

\centerline{Abstract}
We study open and closed string interactions in the Type IIB plane wave
background using open+closed string field theory. We reproduce all string
amplitudes from the dual ${\cal N}=2$ Sp(N) gauge theory by computing
matrix elements of the dilatation operator. A direct diagrammatic
correspondence is found between string theory and gauge theory Feynman
diagrams. The prefactor and Neumann matrices of open+closed string field
theory are separately realized in terms of gauge theory quantities.

\Date{5/2003}

\newsec{Introduction}


Berenstein, Maldacena and Nastase (BMN) \BerensteinJQ\ have recently
found a
particular limit of the AdS/CFT correspondence, in which the {\it free}
string
spectrum in the monochromatic gravitational plane wave of
\BlauNE\BlauDY\ is captured by a subset of operators
in ${\cal N}=4$ SYM (BMN operators). Further tests of the duality were
performed in \GrossSU\ and in \SantambrogioSB,  where the exact free
string
spectrum was derived from purely gauge theory considerations.
Two remarkable ingredients have
been important in
establishing this connection. On the one hand the string worldsheet
theory in the plane wave background \BlauNE\BlauDY\ can be solved
\MetsaevBJ\MetsaevRE\ in
the light-cone
gauge to all orders in $\alpha^\prime$, allowing for the  analysis of
the {\it
free} string spectrum. Perhaps more surprising is the existence
     of a regime in which string theory and the BMN sector of gauge
theory are both perturbative and can be independently  reliably
computed. This remarkable fact allows   the duality to be tested
in this regime, and if quantitative agreement is found,
    it gives us confidence when extrapolating to
     the regime in which perturbation theory breaks down and
we have to rely on the dual description for computation.

Understanding the correspondence in the presence of string
interactions has been more subtle. The  duality
between {\it interacting} string theory
in the plane wave background and gauge theory has recently been
formulated and tested in \GrossMH\GomisWI\GomisKJ\
(see also \PearsonZS\
for a
complementary description using the string bit model
\VerlindeIG\ZhouMI\VamanKA ). The physical principle determining the
correspondence at the interacting level is to identify the {\it full}
interacting Hamiltonian of string theory with that of gauge
theory,  which is the basic premise
of such a holographic correspondence.
      Therefore, the holographic map reads
\eqn\identifiint{{1\over\mu}H=\Delta-J,}
where $H$ is the string Hamiltonian, including all string
corrections,  $\Delta$ is the dilatation operator\foot{In radial
quantization on ${\bf R}^4$, $\Delta$ is the Hamiltonian.}
    of ${\cal N}=4$ SYM,
including  all non-planar corrections and $J$ is the amount of
$U(1)_R$ charge. For other work on   string interactions,
see\foot{In the conclusions we will comment on our understanding of
the invalidity of  other existing
proposals.}
\nref\KristjansenBB{
C.~Kristjansen, J.~Plefka, G.~W.~Semenoff and M.~Staudacher,
``A new double-scaling limit of N = 4 super Yang-Mills theory and
PP-wave  strings,''
Nucl.\ Phys.\ B {\bf 643}, 3 (2002)
[arXiv:hep-th/0205033].
}
\nref\ConstableHW{
N.~R.~Constable, D.~Z.~Freedman, M.~Headrick, S.~Minwalla, L.~Motl,
A.~Postnikov and W.~Skiba,
``PP-wave string interactions from perturbative Yang-Mills theory,''
JHEP {\bf 0207}, 017 (2002)
[arXiv:hep-th/0205089].
}
\nref\BeisertBB{
N.~Beisert, C.~Kristjansen, J.~Plefka, G.~W.~Semenoff
and M.~Staudacher,
``BMN correlators and operator mixing
in N = 4 super Yang-Mills theory,''
arXiv:hep-th/0208178.
}
\nref\KiemXN{
Y.~j.~Kiem, Y.~b.~Kim, S.~m.~Lee and J.~m.~Park,
``pp-wave / Yang-Mills correspondence: An explicit check,''
Nucl.\ Phys.\ B {\bf 642}, 389 (2002)
[arXiv:hep-th/0205279].
}
\nref\HuangWF{
M.~x.~Huang,
``Three point functions of N = 4 super Yang Mills from light cone
string  field theory in pp-wave,''
Phys.\ Lett.\ B {\bf 542}, 255 (2002)
[arXiv:hep-th/0205311].
}
\nref\ChuPD{
C.~S.~Chu, V.~V.~Khoze and G.~Travaglini,
``Three-point functions in N = 4 Yang-Mills theory and pp-waves,''
JHEP {\bf 0206}, 011 (2002)
[arXiv:hep-th/0206005].
}
\nref\LeeRM{
P.~Lee, S.~Moriyama and J.~w.~Park,
``Cubic interactions in pp-wave light cone string field theory,''
Phys.\ Rev.\ D {\bf 66}, 085021 (2002)
[arXiv:hep-th/0206065].
}
\nref\ChuQJ{
C.~S.~Chu, V.~V.~Khoze and G.~Travaglini,
``pp-wave string interactions from n-point correlators of BMN
operators,''
JHEP {\bf 0209}, 054 (2002)
[arXiv:hep-th/0206167].
}
\nref\KlebanovMP{
I.~R.~Klebanov, M.~Spradlin and A.~Volovich,
``New effects in gauge theory from pp-wave superstrings,''
Phys.\ Lett.\ B {\bf 548}, 111 (2002)
[arXiv:hep-th/0206221].
}
\nref\HuangYT{
M.~x.~Huang,
``String interactions in pp-wave from N = 4 super Yang Mills,''
Phys.\ Rev.\ D {\bf 66}, 105002 (2002)
[arXiv:hep-th/0206248].
}
\nref\ParnachevKK{
A.~Parnachev and A.~V.~Ryzhov,
``Strings in the near plane wave background and AdS/CFT,''
JHEP {\bf 0210}, 066 (2002)
[arXiv:hep-th/0208010].
}
\nref\GursoyYY{
U.~Gursoy,
``Vector operators in the BMN correspondence,''
arXiv:hep-th/0208041.
}
\nref\ChuEU{
C.~S.~Chu, V.~V.~Khoze, M.~Petrini, R.~Russo and A.~Tanzini,
``A note on string interaction on the pp-wave background,''
arXiv:hep-th/0208148.
}
\nref\LeeVZ{
P.~Lee, S.~Moriyama and J.~w.~Park,
Phys.\ Rev.\ D {\bf 67}, 086001 (2003)
[arXiv:hep-th/0209011].
}
\nref\deMelloKochNQ{
R.~de Mello Koch, A.~Jevicki and J.~P.~Rodrigues,
``Collective string field theory of matrix models in the BMN limit,''
arXiv:hep-th/0209155.
}
\nref\DobashiAR{
S.~Dobashi, H.~Shimada and T.~Yoneya,
``Holographic reformulation of string theory
on AdS(5) x S**5 background  in the PP-wave limit,''
arXiv:hep-th/0209251.
}
\nref\JanikBD{
R.~A.~Janik,
``BMN operators and string field theory,''
Phys.\ Lett.\ B {\bf 549}, 237 (2002)
[arXiv:hep-th/0209263].
}
\nref\BeisertTN{
N.~Beisert,
``BMN operators and superconformal symmetry,''
arXiv:hep-th/0211032.
}
\nref\ChuWJ{
C.~S.~Chu, M.~Petrini, R.~Russo and A.~Tanzini,
``String interactions and discrete symmetries of the pp-wave
background,''
arXiv:hep-th/0211188.
}
\nref\HeZU{
Y.~H.~He, J.~H.~Schwarz, M.~Spradlin and A.~Volovich,
``Explicit formulas for Neumann coefficients in the plane-wave
geometry,''
arXiv:hep-th/0211198.
}
\nref\RoibanXR{
R.~Roiban, M.~Spradlin and A.~Volovich,
``On light-cone SFT contact terms in a plane wave,''
arXiv:hep-th/0211220.
}
\nref\KiemPB{
Y.~j.~Kiem, Y.~b.~Kim, J.~Park and C.~Ryou,
``Chiral primary cubic interactions from pp-wave supergravity,''
arXiv:hep-th/0211217.
}
\nref\GursoyFJ{
U.~Gursoy,
``Predictions for pp-wave string amplitudes from perturbative SYM,''
arXiv:hep-th/0212118.
}
\nref\CorleyNY{
S.~Corley and S.~Ramgoolam,
``Strings on plane waves, super-Yang Mills in four dimensions, quantum  groups at roots of one,''
arXiv:hep-th/0212166.
}
\nref\MinahanVE{
J.~A.~Minahan and K.~Zarembo,
``The Bethe-ansatz for N = 4 super Yang-Mills,''
arXiv:hep-th/0212208.
}
\nref\BeisertFF{
N.~Beisert, C.~Kristjansen, J.~Plefka and M.~Staudacher,
``BMN gauge theory as a quantum mechanical system,''
arXiv:hep-th/0212269.
}
\nref\ChuQD{
C.~S.~Chu and V.~V.~Khoze,
``Correspondence between the 3-point BMN correlators and the 3-string
vertex on the pp-wave,''
arXiv:hep-th/0301036.
}
\nref\KloseTW{
T.~Klose,
``Conformal dimensions of two-derivative BMN operators,''
arXiv:hep-th/0301150.
}
\nref\GeorgiouAA{
G.~Georgiou and V.~V.~Khoze,
``BMN operators with three scalar impurites and
the vertex-correlator  duality in pp-wave,''
JHEP {\bf 0304}, 015 (2003)
[arXiv:hep-th/0302064].
}
\nref\SpradlinBW{
M.~Spradlin and A.~Volovich,
``Note on plane wave quantum mechanics,''
arXiv:hep-th/0303220.
}
\nref\DiVecchiaYP{
P.~Di Vecchia, J.~L.~Petersen, M.~Petrini, R.~Russo and A.~Tanzini,
``The 3-string vertex and the AdS/CFT duality in the pp-wave limit,''
arXiv:hep-th/0304025.
}
\nref\YoneyaMU{
T.~Yoneya,
``What is holography in the plane-wave limit
    of AdS(5)/SYM(4)  correspondence?,''
arXiv:hep-th/0304183.
}
\nref\PankiewiczKJ{
A.~Pankiewicz,
``An alternative formulation of light-cone string
    field theory on the  plane wave,''
arXiv:hep-th/0304232.
}
\nref\FreedmanBH{
D.~Z.~Freedman and U.~Gursoy,
``Instability and degeneracy in the BMN correspondence,''
arXiv:hep-th/0305016.
}
\nref\deMelloKochPV{
R.~de Mello Koch, A.~Donos, A.~Jevicki and J.~P.~Rodrigues,
``Derivation of string field theory from the large N BMN limit,''
arXiv:hep-th/0305042.
}
\hskip-58pt$\!\!\!\!\!\!\!\!\!\!\!\!\!\!\!\!\!\!\!\!\!\!\!\!
\!\!\!\!\!\!$\KristjansenBB -\deMelloKochPV .

In \GomisWI, a conjectured exact
mapping between gauge theory and string theory states was given
to all orders in
perturbation theory.
A unique basis of states $|\tilde{O}_B\ra$ in gauge theory was
identified and proposed to be the correct one to use when comparing
with string theory Hamiltonian matrix elements of arbitrary string
states
$|s_A\ra$.  The conjectured correspondence between  free string states -- 
eigenstates of the free Hamiltonian --  
and gauge theory states to all orders in perturbation theory is
therefore:
\eqn\mapping{
|s_A\ra\leftrightarrow |\tilde{O}_B\ra.}
     The basis of gauge theory states $|\tilde{O}_B\ra$ -- henceforth
     named the  string
basis\foot{When $g_2=0$ these operators
    reduce to the BMN operators in
\BerensteinJQ, which are dual to  unperturbed string states.}
    -- is implicitly
defined within gauge theory and makes no reference to string
theory.  The precise proposal
for the duality is given by
\eqn\interact{
{1\over \mu}
\la s_A|H|s_B\ra
=\la\tilde{O}_A|(\Delta-
J)|\tilde{O}_B\ra=n\delta_{AB}+\tilde{\Gamma}_{AB},}
where $\tilde{\Gamma}_{AB}$ is the
matrix of anomalous dimensions in
the string basis, $|s_A\ra$ are the string Fock space states
     and $n$ is the number of impurities/oscillators carried by the
state.
    Therefore, string interactions
are captured by non-planar corrections to the matrix of anomalous
dimensions in the string basis.
This holographic map applies to
the duality for any interaction and to
all orders in both\foot{This parameter identification is valid for
rank $N$ $Sp(N)$ gauge theory. Here $g^2=4\pi g_s$.}
    $g_2\equiv {J^2\over 2N}=4\pi g_s(\mu
p^+\alpha^\prime)^2$ and
$\lambda^\prime\equiv{g^2(2N)\over J^2}={1\over (\mu 
p^+\alpha^\prime)^2}$.

Physically,
the string basis in \GomisWI\ is the
unique basis of states in gauge theory which is
{\it orthonormal} and completely {\it determined} in terms of the inner
product
$G_{AB}$ of  BMN
     operators\foot{We recall that $G_{AB}$
    and  ${\Gamma_{AB}}$ can be
extracted from the matrix of two-point functions
\eqn\teopoint{|2\pi x|^{2\Delta_0}\la O_A\bar{O}_B\ra
=G_{AB}+\Gamma_{AB}\ln(x^2\Lambda^2)^{-1},}
where $O_A$ denote the BMN operators.}.
 The change of basis
transformation
from the
string basis to the BMN basis is the unique  real and symmetric
  matrix which
orthonormalizes the inner product.
In \GomisWI, an explicit
algorithm for constructing the string basis in terms of BMN states to
any desired order in
$g_2$ was given and explicit formulas were given for the first few
terms. Once the basis is identified, one can compute the matrix of
anomalous dimensions in this basis, $\tilde{\Gamma}_{AB}$, and relate it
to purely gauge theory quantities computable from the two-point function
of BMN
operators, that is in terms of $G_{AB}$ and $\Gamma_{AB}$. One can
then test whether these matrix elements agree with the string
Hamiltonian matrix elements \interact. The algorithm in \GomisWI\ yields
the
following predictions\foot{In this paper since we consider a gauge
theory with  $Sp(N)$ gauge symmetry and fundamental matter and the
expansion parameter is $\sqrt{g_2}$, which corresponds to  the open
string  coupling constant. These formulas can be obtained
from \GomisWI\ by replacing
the index $s$ in $M^{(s)}$ by $s/2$. See around $(2.12)$
for a discussion of the expansion parameters.}
\eqn\anombasis{\eqalign{
\tilde\Gamma^{(0)}&=\Gamma^{(0)},\cr
\tilde\Gamma^{({1\over 2})}&=\Gamma^{({1\over 2})}
-{1\over 2}\{G^{({1\over 2})},\Gamma^{(0)}\},\cr
\tilde\Gamma^{(1)}&=\Gamma^{(1)}-{1\over2}\{G^{(1)},\Gamma^{(0)}\}
-{1\over2}\{G^{({1\over 2})},\Gamma^{({1\over 2})}\}
+{3\over8}\bigl\{\bigl(G^{({1\over 2})}\bigr)^2,\Gamma^{(0)}\bigr\}
+{1\over4} G^{({1\over 2})}\Gamma^{(0)}G^{({1\over 2})},\cr
\tilde\Gamma^{({3\over 2})}&=\Gamma^{({3\over 2})}
-{1\over 2}\{G^{({1\over 2})},\Gamma^{(1)}\}
-{1\over 2}\{G^{(1)},\Gamma^{({1\over 2})}\}
+{3\over 8}\bigl\{\bigl(G^{({1\over 2})}\bigr)^2,
\Gamma^{({1\over 2})}\bigr\}
+{1\over 4}G^{({1\over 2})}\Gamma^{(0)}G^{({1\over 2})}\cr
&\quad-{1\over 2}\{G^{({3\over 2})},\Gamma^{(0)}\}
+{3\over 8}\bigl\{\bigl\{G^{(1)},G^{({1\over 2})}\bigr\},
\Gamma^{(0)}\bigr\}
-{1\over 16}\bigl\{\bigl(G^{({1\over 2})}\bigr)^3,\Gamma^{(0)}\bigr\}\cr
&\quad+{1\over 4}G^{(1)}\Gamma^{(0)}G^{({1\over 2})}
+{1\over 4}G^{({1\over 2})}\Gamma^{(0)}G^{(1)}
-{3\over 16}G^{({1\over 2})}\Gamma^{(0)}\bigl(G^{({1\over 2})}\bigr)^2
-{3\over 16}\bigl(G^{({1\over 2})}\bigr)^2
\Gamma^{(0)}G^{({1\over 2})},\cr
&\ \vdots}}
where $M^{(s)}$, with $M=\tilde\Gamma,\ \Gamma$ or $G$, is the $g_2^s$
term in the expansion of
$M=M^{(0)}+g_2^{{1/2}}M^{({1\over 2})}+g_2M^{(1)}+g_2^{3/2}M^{({3\over
2})}+\cdots$. As in any duality, the two
sides of the duality can be independently computed without any
reference to the dual theory. Comparison of the two independent
calculations and using \interact\ determines whether the proposed
holographic map is
correct. In particular, the proposal in \GomisWI\ for
the gauge theory basis dual
to string states can be falsified by comparing with string theory
computations.

     The validity of the holographic map \identifiint\ and of the
basis of gauge theory states \anombasis\  dual to string states
has been confirmed in
\PearsonZS\GomisWI\GrossMH\
    for two different impurities and extended
to arbitrary impurities in \GomisKJ.
Moreover, in  \GomisKJ\ a direct
diagrammatic correspondence was
found between string theory and gauge
theory
Feynman diagrams. Both the prefactor
and {\it all} Neumann matrices of interacting
string
theory -- computed in \HeZU -- were
    independently transcribed in terms of pure
gauge theory data to leading order in $\lambda^\prime$.

In this paper, we continue the investigation of string interactions
in the plane-wave background by adding open strings to
the system.  We will consider a  four
dimensional ${\cal N}=2$ gauge theory with $Sp(N)$ gauge symmetry and
matter in the $\anti\oplus 4\fund$ representations of $Sp(N)$. This
gauge theory was shown \BerensteinZW\ to describe the unoriented open
and closed {\it free}
string spectrum of an orientifold projection of the maximally
supersymmetric plane with $4$ D7-branes on top of an O$7^-$-plane.
Having open strings greatly augments the types of string
interactions that can occur, which are captured by full open+closed
(super)string field theory \GreenFU.
    In unoriented open+closed string field theory
there are seven different interactions that can occur and each one of
them is described by an interaction term in the string
Hamiltonian $H$. The holographic map \interact\ naturally extends to all
of these interactions to all orders in $\lambda^\prime$. This open+closed
system is a very rich one in which the holographic
map \identifiint\  and our mapping of states \anombasis\ can be
thoroughly tested. For work on this open+closed system, see
\BerensteinSA\ImamuraWZ .

Here we study in detail some of the possible open+closed interactions using
string
field theory and gauge theory. We find, using the  universal
holographic map
\interact\ and the universal string basis \anombasis, that we precisely
reproduce all string amplitudes for states with arbitrary number of
scalar
impurities to leading order in $\lambda^\prime$. Moreover, as in
\GomisKJ, we find  for a given string interaction that there is a
one-to-one correspondence between string theory and gauge theory
Feynman diagrams. Furthermore,
    we can separately transcribe the prefactor
and  the Neumann matrices
for all string interactions purely in terms of gauge theory
data to leading order in $\lambda^\prime$.

  We show that the two different prefactors \GreenFU\ of
open+closed string field theory
are described by two different quartic interactions
in the gauge theory, realizing respectively the ``joining-splitting''
and ``exchange'' interaction of strings.
The Neumann matrices to leading order in the large $\mu$ expansion
for any of the seven
interactions
are described in gauge theory by the sum over all free
contractions of the corresponding  impurity in the corresponding
gauge theory two-point function.

The rest of the paper is organized as follows. In section $2$, we
describe the  BMN operators dual to free open and closed strings. We
summarize the string theory and gauge theory ingredients that are
required to perform the detailed comparison, and outline
the relation of the  prefactor and Neumann matrices in
string theory to gauge theory quantities, which has common
features for all interactions. In section $3$, we perform the detailed
string theory computation of the cubic open string transition
amplitudes. The corresponding gauge theory
computation and exact agreement is exhibited in section 4.
Section $5$ contains the
string theory computation of the amplitude for an open string to
become a closed string, which is followed by the corresponding gauge
theory computation in section 6.
We finish with some conclusions. The Appendices contain formulas and
derivations  
needed in the main text.

\newsec{Duality with Open and Closed Strings}

A simple way of adding open strings to the gravity side of the AdS/CFT
correspondence is to consider the near horizon limit of a ``large''
collection of D3-branes probing  a ``small'' set of other
     D-branes. When taking the near
horizon limit, the D3-branes become the familiar $AdS_5\times S^5$
background while the ``small'' number of extra branes can be treated as
probe
branes in the $AdS_5\times S^5$
geometry. A simple realization of this idea is to take the near
horizon limit of $N$ D3-branes probing an O$7^{-}$-plane together with
the accompanying $4$ D7-branes required to cancel tadpoles. In the
near horizon limit one gets \AharonyXZ\
     an
O$7^{-}$-plane with  D7-branes whose worldvolume is $AdS_5\times S^3$
inside $AdS_5\times S^5/Z_2$. Therefore, in the bulk one has closed
and open strings. In particular, at low energies one has supergravity
together with an $SO(8)$ gauge theory living on the D7-branes.
The dual gauge theory \DouglasJS\
is a $D=4$ ${\cal
N}=2$ $Sp(N)$ gauge theory with hypermultiplets in the
$\anti\oplus 4\fund$ representations of $Sp(N)$.
     The AdS/CFT duality
states that this gauge theory captures the closed and open string
physics inside $AdS_5\times S^5/Z_2$.

\noindent
$\bullet$ {\it Free String Limit}

Progress towards capturing {\it stringy} physics can be made by taking
the plane wave limit. By taking the boosted great circle inside
$S^5/Z_2$ to
be along a worldvolume direction of the D7-branes, one
obtains \BerensteinZW\ an orientifold projection
of the familiar Type IIB plane wave
\BlauNE\BlauDY\  together with an O$7^-$-plane and $4$ D7-branes sitting
at
$x_7=x_8=0$  in the transverse ${\bf R}^8$ directions of the plane
wave\foot{For other realizations of open strings in plane wave
background, see \LeeCU\Sken\Bala\NaculichFH.}. The D7-branes break the $SO(8)$
symmetry acting on the transverse
${\bf R}^8$ down to $SO(6)\times U(1)_{78}$ which acts on ${\bf
R}^6\times {\bf
R}_{78}^2$. Moreover, the five-form flux further breaks the symmetry
down to $SO(4)\times U(1)_{56}\times U(1)_{78}$ which acts linearly
on ${\bf
R}^4\times{\bf
R}_{56}^2\times {\bf
R}_{78}^2$.  Therefore, the transverse
$SO(8)$ vector index $I$ decomposes into
$I\rightarrow (x^\mu,z^\prime,\bar z^\prime, w, \bar w)$, where
$\mu=1,\ldots,4,\ z^\prime={1\over \sqrt{2}}(x^5+ix^6)$ and
$w={1\over \sqrt{2}}(x^7+ix^8)$. Therefore, the N(eumann) directions
correspond to $N\in\{x^\mu,z^\prime,\bar z^\prime\}$ and the D(irichlet)
directions to $D\in\{w,\bar w\}$.

    The free open string spectrum on the D7-branes was
reproduced  from
gauge theory computations to leading order in the $\lambda^\prime$
expansion
\BerensteinZW\ImamuraWZ . In this paper we will compute string
interactions among open and closed strings from gauge theory using the
proposal
\interact\anombasis\  and show that they exactly
agree with the string field theory calculation of the interactions.

In an ${\cal N}=1$ language, the various ${\cal N}=2$ multiplets can
be decomposed as:
\eqn\decomp{\eqalign{
\hbox{vector multiplet}&\rightarrow (V,W),\cr
\anti \hbox{hypermultiplet}&\rightarrow (Z,Z^\prime),\cr
4\fund \hbox{hypermultiplets}&\rightarrow ({\tilde q}_A,q_A)\quad
A=1,\ldots,4,}}
where $V$ is an ${\cal N}=1$ vector multiplet while
$W,Z,Z^\prime,{\tilde q}_A$ and $q_A$ are ${\cal N}=1$ chiral multiplets
transforming in the appropriate representations of $Sp(N)$.
The gauge theory has $SU(2)_R \times U(1)_R$ $\cal R$-symmetry
     and $SU(2)_L \times SO(8)$ global symmetry. We  make manifest the
$SO(8)$ global symmetry of the gauge theory
by taking linear combinations of the quarks and forming an $SO(8)$ vector
     $Q^i\; (i=1,\ldots,8)$.


\noindent

The closed string vacuum state is identified as before
with\foot{$\Omega^{ab}$ is
the invariant antisymmetric matrix which raises and lowers $Sp(N)$
indices.}
\eqn\BMNclovac{O^J_{\rm vac}
={1\over \sqrt{2J(2N)^{J}}}\hbox{Tr}[(Z\Omega)^J]
\quad\Leftrightarrow\quad |\hbox{vac}\ra\qquad \Delta-J=0,}
and the open string vacuum state is given by the unique chiral primary
operator with two quarks:
\eqn\BMNopenvacb{O^{J;ij}_{\rm vac}
={1\over \sqrt{(2N)^{J}}} Q^i\Omega(Z\Omega)^{J-1}Q^j
\quad\Leftrightarrow\quad |\hbox{vac};ij\ra\qquad \Delta -J=1.}
An actual open string state is obtained by specifying a Chan-Paton
wavefunction $\lambda$, given by a hermitian matrix
$\lambda_{ij}$. The correspondence is therefore:
\eqn\BMNopenvac{O^{J}_{\rm vac}
={1\over \sqrt{(2N)^{J}}} \lambda_{ij}Q^i\Omega (Z\Omega)^{J-1}Q^j
\quad\Leftrightarrow\quad \lambda_{ij}|\hbox{vac};ij\ra\qquad \Delta
-J=1,}
where $\lambda$ is normalized such that
    $\hbox{Tr}(\lambda^a \lambda^b) = \delta^{ab}/2$. There are two
differences in the normalization of BMN operators compared with a $U(N)$
gauge theory. First, each
index loop contributes a factor of $2N$ instead of $N$ since the
fundamental
representation of $Sp(N)$ is $2N$-dimensional. In addition,
fields in $Sp(N)$ theory are ``unoriented" and one
    can contract two fields in two ways,
with an ordinary propagator or with a twisted propagator.
In the planar limit, we are forced to use the same propagator for
all fields. Using the twisted propagator for all fields is equivalent
to transposing one operator in a two-point function and replacing
the twisted propagator with the ordinary one.
These two possibilities give us an extra combinatoric
factor of $2$ in \BMNclovac. For \BMNopenvac, this factor of $2$ is
absorbed  in the normalization of the Chan-Paton wave functions.

When $g_2=0$, i.e., in the planar limit, the BMN
operators describing free strings are obtained by diagonalizing the
dilatation operator $\Delta$ to leading order in
$\lambda^\prime$ \BerensteinZW\ImamuraWZ. This procedure can be carried 
both for operators
describing closed and open strings, which correspond to operators
with a trace and operators capped by fundamental quarks respectively.
     One must write down all operators
in a given charge sector and diagonalize the matrix of two-point
functions. The eigenvectors of this matrix yield the gauge theory
realization -- the BMN operators -- of free open and closed strings. The
operators with lowest number of scalar impurities\foot{The other 4
bosonic impurities correspond to $D_\mu\cdot$ insertions.} and the
corresponding string states that we will need in this paper  are
respectively
given by ($X,Y=N\ \hbox{or}\ D$, where $N\in \{Z^\prime,\bar{Z^\prime}\}$
and $D\in
\{W,\bar{W}\}$):

\noindent
$\bullet$ Closed Strings:\foot{$\alpha_n$, $a_n$ and $a_{-n}$
oscillators are associated with
exponential Fourier modes, $\cos$ modes and $\sin$ modes
respectively. As in \GomisKJ\ the overall phase of the string states
dual to BMN operators is $i^s$, where $s$ is the number of
impurities carried by the string state.}
\eqn\BMNclo{\eqalign{
&O^J_{n,(X,Y)}\hskip-3pt=\hskip-3pt{1\over 
\sqrt{J(2N)^{J+2}}}\hskip-2pt\left(\hskip-1pt\sum_{l=0}^J
e^{2\pi in\over J}
\hbox{Tr}\Big[(Y\Omega)(Z\Omega)^l(X\Omega)(Z\Omega)
^{J-l}\Big]\hskip-2pt-\hskip-2pt\delta_{X,{\bar
Y}}\hbox{Tr}\Big[({\bar Z}\Omega)
(Z\Omega)^{J+1}\hskip-2pt\Big]\hskip-1pt\right)\cr
&\qquad\qquad\qquad\qquad\Leftrightarrow\quad
|XY,n\ra=-{1\over\sqrt{2}}(\alpha_{n}^{X\dagger}
\alpha_{-n}^{Y\dagger}+
(-1)^{\#_D}\alpha_{-n}^{X\dagger}
\alpha_{n}^{Y\dagger})|\hbox{vac}\ra,}}
where $\#_D$ denotes the number of Dirichlet impurities.
Notice that we don't have the aforementioned factor of
$1/\sqrt{2}$ for  nonzero $n$ because the phase
prevents the overlap with twisted propagators. Nevertheless,
the zero momentum operators should have an extra factor of $1/\sqrt{2}$.

\noindent
$\bullet$ Open Strings:

\noindent
(1) Neumann Impurities\foot{When we insert a $\bar Z^\prime$ impurity,
there
is a boundary term with an antiquark. However, this term is subleading
in $1/J$
     and does not contribute throughout this paper.}:
\eqn\BMNopen{\eqalign{
&O^{J}_{n,N}={1\over\sqrt{J(2N)^{J+1}}}
\sum_{l=0}^{J-1}\sqrt{2}\cos{\left({\pi n\over
J}\right)}\lambda_{pq}Q^p\Omega (Z\Omega )^l
(N\Omega )(Z\Omega )^{J-1-l}Q^q \cr
&\qquad\qquad\qquad\qquad\Leftrightarrow\quad
|N,n\ra=i\lambda_{pq}a_{n}^{N\dagger}|\hbox{vac};pq\ra.}}
\noindent
(2) Dirichlet Impurities:
\eqn\BMNopenb{\eqalign{
&O^{J}_{n,D}={1\over\sqrt{J(2N)^{J+1}}}
\sum_{l=0}^{J-1}
\sqrt{2}\sin{\left({\pi n\over J}\right)}\lambda_{pq}
Q^p(Z\Omega)^l(D\Omega)(Z\Omega)^{J-1-l}Q^q\cr
&\qquad\qquad\qquad\qquad\Leftrightarrow\quad
|D,n\ra=i\lambda_{pq}a_{-n}^{D\dagger}|\hbox{vac};pq\ra.}}

\noindent
(3) Two-string States:
\eqn\BMNmulv{\eqalign{
T^{y,J}_{\rm vac}
=\lambda^1_{ij}\lambda^2_{kl}:O^{y\cdot J;ij}_{\rm
vac}O^{(1-y)\cdot
J;kl}_{\rm vac}:
&\Leftrightarrow
|{\rm vac},y\ra\ra=\lambda^1_{ij}|{\rm vac},y;ij\ra\otimes
\lambda^2_{kl}
|{\rm vac},1-y;kl\ra,\cr
T^{y,J}_{n,X,(1)}
=\lambda^1_{ij}\lambda^2_{kl}:O^{y\cdot J;ij}_{n,X}O^{(1-y)\cdot
J,kl}_{\rm vac}:
&\Leftrightarrow
|(X,n)_1,y\ra\ra\hskip-1pt=\hskip-1pt\lambda^1_{ij}|(X,n),y;ij\ra
\hskip-2pt\otimes \hskip-2pt\lambda^2_{kl}|{\rm
vac},1\hskip-2pt-\hskip-2pt y;kl\ra,\cr
T^{y,J}_{n,X,(2)}
=\lambda^1_{ij}\lambda^2_{kl}:O^{y\cdot
J,ij}_{\rm vac}O^{(1-y)\cdot J;kl}_{n,X}:
&\Leftrightarrow
|(X,n)_2,y\ra\ra\hskip-1pt=\hskip-1pt\lambda^1_{ij}|{\rm vac},y;ij\ra
\hskip-2pt\otimes\hskip-2pt
\lambda^2_{kl} |(X,n),1\hskip-2pt-\hskip-2pty;kl\ra,}}
where $0<y<1$ is the fraction of the total longitudinal 
momentum carried by the
first string in the two-string state and the $\lambda$'s are the
Chan-Paton wavefunctions of the open string states.

A very simple selection rule can be established for the BMN operators
dual to open and closed strings using the symmetry properties\foot{The
matrix $W_{ab}$ is symmetric while $Z_{ab},Z_{ab}^\prime$ are
antisymmetric.}of the
fields \BerensteinZW. One can show that they realize
           the action of the orientifold group\foot{For closed string
oscillators the action of the orientifold is $\alpha^N_n\leftrightarrow
\alpha^N_{-n}$ and  $\alpha^D_n\leftrightarrow
-\alpha^D_{-n}$ while for open strings it is given by
   $a_n^I\rightarrow (-1)^n a_n^I$.}
   on the dual free string
states. For closed string BMN operators only the symmetric or
antisymmetric part of the operator under the exchange of the
two impurities is non-zero, encoding the proper combination of closed
string states surviving the orientifold projection.
The  open string
BMN operators have different transformations properties under the
exchange of the quarks (endpoints of the string)
depending on the the amount of worldsheet momentum which they carry
\eqn\parity{\eqalign{
n\ \hbox{even}\quad&\rightarrow\quad \lambda^\T=-\lambda,\cr
n\ \hbox{odd}\quad&\rightarrow\quad \lambda^\T=\lambda,}}
so that states with $n$ even transform in the antisymmetric (adjoint)
representation
of $SO(8)$ while states with $n$ odd transform in the symmetric
representation of $SO(8)$. This is precisely the transformation
properties of unoriented open strings on a D7-brane.

    In general, we can add many impurities, and operators
are defined in a similar way to \GomisKJ.
    Then, the Chan-Paton matrix has the symmetry property
\eqn\CPsymm{ \lambda^\T =
    (-1)^{1+\sum_i n_i} \lambda,}
where the sum is over each impurity
    and $n_i$ is its worldsheet momentum.

\vfill\eject

\noindent
$\bullet$ {\it Interacting Open and Closed Strings}

When $g_2\neq 0$ open and closed strings can interact, and on the gauge
theory side, non-planar corrections become important. The
holographic map \interact\ dictates  that the matrix elements of the
string Hamiltonian between arbitrary string states (open or closed)
to ${\cal O}(g_2^s)$ are captured by
${\cal O}(g_2^s)$ non-planar corrections to the matrix of anomalous
dimensions of the dual gauge theory states.

\noindent
$1)$ String Theory Computation:

Light-cone (super) string field theory \GreenTC\GreenHW\GreenFU\GreenQU\
provides a precise formalism in which
to compute string interactions using old fashioned Hamiltonian
perturbation theory.  In unoriented open and closed string theory,
which is the one considered in this paper, there are seven types of
basic interactions that can occur between strings. Each interaction
is characterized by the number of open and closed string states in the
final states, and it is captured by a particular interaction term in
the string Hamiltonian. $\sqrt{g_2}$ and $g_2$
denote the open and closed string coupling constants respectively.
Schematically, the string Hamiltonian is given
by\foot{Here we omit the infamous higher order contact terms.}
\eqn\stringHa{
H=H_{cc}+H_{oo}+\sqrt{g_2}(H_{o\leftrightarrow
oo}+H_{o\leftrightarrow c})+g_2(H_{c\leftrightarrow
cc}+H_{o\leftrightarrow oc}+H_{oo\leftrightarrow
oo}+H_{o\leftrightarrow o}+H_{c\leftrightarrow c}),}
where $H_{cc}, H_{oo}$ is the free closed and open string Hamiltonian and
the rest are the various interaction terms.
In oriented open and closed string theory, the last two terms in
\stringHa\ are absent since they describe  a self-interacting process,
    which is orientation non-preserving.

Each interaction term in the Hamiltonian can be computed following the
seminal flat space analysis in \GreenFU . Each term has two basic
ingredients. One is purely geometrical and encodes the geometrical
gluing of the strings involved in the interaction. In addition,
  it is summarized by a
Mandelstam diagram. This piece is usually referred to as the overlap.
    The second contribution, called the prefactor,
encodes the behaviour of the worldsheet at the interaction
point. Despite the fact that the external states can be different, near
the interaction point, there are only two basic types of ``gluing'' of
strings. One is a ``joining-splitting" type of interaction corresponding
to interactions at ${\cal O}(\sqrt{g_2})$ and the other one is an
``exchange'' type of interaction corresponding
to interactions at ${\cal O}(g_2)$. The basic intuition that there are
only two different prefactors was beautifully demonstrated in
\GreenFU\ by a careful analysis of the supersymmetry
algebra. Some work  on string field theory in  the plane wave
background
can be found in
\SpradlinAR\SpradlinRV\SchwarzBC\PankiewiczGS\PankiewiczTG
\HeZU\PankiewiczKJ\DiVecchiaYP\ChandrasekharFQ\StefanskiZC.

The overlap function $|V\ra$ is represented by a squeezed state and
encodes the continuity of all worldsheet fields in the interaction
diagram. The quantity specifying $|V\ra$ are the Neumann matrices. When
comparing to perturbative gauge theory, we need the large $\mu$
expression for the Neumann matrices. One can prove for each of the
seven interaction terms in \stringHa\ that the large $\mu$ Neumann
matrices reduce
precisely to the Fourier overlap on the corresponding worldsheet diagram
which
allows one to rewrite the oscillators of the outgoing strings in the
diagram in terms of the oscillators of the incoming strings. This
general result is crucial in establishing the equivalence with gauge
theory and is proven in Appendix B by demanding continuity of the
worldsheet fields in  the interaction diagram.

The computation is conveniently performed
\GomisKJ\  by introducing Feynman rules
for each term in the interaction Hamiltonian. As we will show in the
upcoming sections equivalence with gauge theory is established diagram
by diagram. Moreover, both the prefactor and the overlap function can
be separately transcribed in terms of gauge theory quantities.

\noindent
$2)$ Gauge Theory Computation:

Having outlined how to perform the string computation for any process,
let us turn at the corresponding (independent) gauge theory
computation. As explained in the introduction, the conjecture is that
the Hamiltonian matrix elements are captured by the matrix of
anomalous dimensions \interact\ of the dual gauge theory states proposed
in
\GomisWI\ and summarized in \anombasis . The expressions in
\anombasis\ are given in terms of
     the two-point function of the BMN
operators given in \BMNclo-\BMNopenb .  There are a few simple
observations we can make about these correlators. First, there is no
free contraction in the two-point function of open and/or closed
operators
at ${\cal O}(\sqrt{g_2})$. Since we normalize each operator such that
     it is unit normalized in the planar limit, we can rewrite
a two-point function as
\eqn\twonorm{
\la O_1\bar O_2\ra|_{\rm free} =
  {\la O^{\rm bare}_1\bar O^{\rm bare}_2\ra|_{\rm free}
    \over \sqrt{\la O^{\rm bare}_1\bar O^{\rm bare}_1\ra
|_{\rm free}^{\rm planar}}
     \sqrt{\la O^{\rm bare}_2\bar O^{\rm bare}_2
\ra|_{\rm free}^{\rm planar}}},}
where $O_i$'s are BMN operators and
  $O^{\rm bare}_i$'s are the corresponding
operators without an $N$-dependent
    normalization factor.
In order to have a non-zero {\it free} contraction,
the number of bulk fields and of quark fields should be preserved
separately. Hence, as long as $N$ counting is concerned,
$\la O^{\rm bare}_1\bar O^{\rm bare}_1\ra
|_{\rm free}^{\rm planar}
= \la O^{\rm bare}_2\bar O^{\rm bare}_2
\ra|_{\rm free}^{\rm planar}$.
Therefore,
\eqn\twonormb{
\la O_1\bar O_2\ra|_{\rm free} =
  {\la O^{\rm bare}_1\bar O^{\rm bare}_2\ra|_{\rm free}
    \over \la O^{\rm bare}_1\bar O^{\rm bare}_1\ra
|_{\rm free}^{\rm planar}}.}
Then,
the familiar 't Hooft counting shows that there cannot arise a
half-integral
power of $N$.
For a {\it free} contraction this implies that we do not get a
half-integral
power of $g_2$.
$\sqrt{g_2}=J/\sqrt{2N}$ corrections arise only from
     a process which does not preserve the  number of quarks
     and this necessitates an interaction vertex involving quarks and bulk
fields.
Consequently, a half-integral power of $g_2$ is always accompanied by
a loop factor of $\lambda^\prime$.
     Therefore, the mixing matrix at  ${\cal
O}(\sqrt{g_2})$ vanishes, that is $G^{({1\over 2})}_{AB}=0$, and to this
order
in the $g_2$ expansion, the string basis of gauge theory states is
identical to the
free BMN basis. Thus, it follows from the proposal in \anombasis\
that the dual description of the $H_{o\leftrightarrow
oo}$ and $H_{o\leftrightarrow c}$ string interactions are identically
given by the matrix of anomalous dimensions in the BMN basis:
\eqn\openreduce{
\tilde\Gamma^{({1\over 2})}=\Gamma^{({1\over 2})}.}

Moreover, the dual description of the ${\cal O}(g_2)$ interactions in
\stringHa\ using \anombasis\ reduces to:
\eqn\truncate{
\tilde\Gamma^{(1)}=\Gamma^{(1)}-{1\over2}\{G^{(1)},\Gamma^{(0)}\}.}
A  simple
consistency check that follows from the result that  $G^{({1\over
2})}_{AB}=0$ is that the formula for $\tilde\Gamma^{(1)}$ in
\anombasis\ reduces to the formula derived in \GomisWI . This fits
nicely with the fact that the formula for the ${\cal O}(g_2)$ cubic
closed string
Hamiltonian is exactly the same as in the oriented string theory, so that
the agreement found in \PearsonZS\GomisWI\GrossMH\GomisKJ\ for closed
string amplitudes still
holds.

The goal is therefore to compute the two point function of BMN
operators at one loop ${\cal O}(\lambda^\prime)$. There are two basic
steps in the computation. One is to insert a loop in the two point
function. After taking into consideration important cancellations
among different terms in the Lagrangian of the gauge theory -- which is
summarized in Appendix A --  there are
two quartic interactions which give non-vanishing contributions.
    These interactions couple the impurities in the operators in the two
point function.
    As we show in the following sections, we find a direct
relation between the two quartic interactions and the two types of
prefactors in string field theory, corresponding respectively to the
``joining-splitting'' interaction and to the ``exchange''
interaction. The identification is given by\foot{For the ``exchange''
interaction we present the identification for the $Z^\prime$ impurity.}:
\eqn\mapinte{\eqalign{
{\cal L}_{\rm int}=
g^2 \bar Q^i\Omega[Z\Omega,Z^\prime\Omega]
\bar Q^i &\ \leftrightarrow
\ \hbox{``joining-splitting'' interaction},\cr
{\cal L}_{\rm int}=-g^2{\rm Tr}\left(
[Z\Omega,Z^\prime\Omega][\bar Z\Omega,
  \bar Z^\prime\Omega]\right)
&\ \leftrightarrow
\ \hbox{``exchange'' interaction}.}}

In computing the effect of the interaction \mapinte\ we must keep
track of the phases that are produced by the interaction. Moreover,
to ${\cal O}(\lambda^\prime)$  the rest of the impurities in the
operators must be freely contracted. The free contraction of each
impurity gives rise to a non-trivial sum over phases due to the
definition of the BMN operators \BMNclo-\BMNmulv, where each impurity has
associated a phase proportional to the amount of worldsheet
momentum that it carries. We show that the
sum over free contractions of an  impurity for a given two point
function dual to a string interaction, is exactly the same as
the numerical value of the Neumann matrix of the corresponding string
diagram in the large $\mu$ limit as shown in Appendix B. Thus, we have
the following 
correspondence:
\eqn\sumphase{
\hbox{sum over free contractions for one impurity}\leftrightarrow
\ \hbox{Neumann matrix}.}
With the ingredients in \mapinte\sumphase\ we do not only find
equivalence diagram by diagram between string theory and gauge theory,
but we can transcribe the constituents of the string Hamiltonian in
purely gauge theory terms.

In the process of comparing string field theory with gauge theory one
must take care of how states are normalized
\ConstableHW\BerensteinSA.
String field theory canonically computes with states which have delta
function
normalization  $\la
s_A|s_B\ra=|p_A^+| \delta(p_A^++p_B^+)=J_A
\delta_{J_A,J_B}$ while the dual gauge theory states $|\tilde{O}_A\ra$
are unit normalized. Moreover, one must also take  into account an
overall delta 
function
conservation
$|p^{+}_{(3)}|\delta(p^{+}_{(1)}+p^{+}_{(2)}+p^{+}_{(3)})=J\delta_{J_1+J_2,
J_3}$
in the
string field theory vertex. With these ingredients it is easy to find
the factor that one must multiply the gauge theory answer before
comparing with string field theory. For example, when comparing
with $H_{o\leftrightarrow oo}, H_{c\leftrightarrow cc}$ one  must
multiply the gauge theory
answer by $\sqrt{Jy(1-y)}$, where $y=-p^{+}_{(1)}/p^{+}_{(3)}=J_1/J$ and
$1-y=-p^{+}_{(2)}/p^{+}_{(3)}=J_2/J$ while there is no factor that
needs to be inserted when comparing with  $H_{o\leftrightarrow c},
H_{c\leftrightarrow c}$.

Before starting the computation we would like to describe which string
interactions can be reliably computed using {\it perturbative} gauge
theory . In the large $\mu$ limit, any string state sits in a
particular energy band, the energy difference of whose states is of
order $\delta
E\sim \mu\lambda^\prime$ \ConstableHW. States in a given band have the
same number
of impurities. States with different number of impurities have a gap
of order  $\delta
E\sim \mu$. Yang-Mills can {\it perturbatively} reproduce string
amplitudes
for which the energy difference between
    the {\it in} and {\it out} states is small (of order  $\delta
E\sim \mu\lambda^\prime$). In string theory we can easily compute
amplitudes between states whose energy difference is large, but those
amplitudes are {\it nonperturbative} in gauge theory.
In particular, for the case of closed
string interactions, only matrix elements between states with the same
number of impurities are perturbative. In the presence of open strings,
things are different. Since the energy of the vacuum open string state
\BMNopenvac\ is $\mu$, demanding the energy difference between {\it in}
and {\it out} states
to be small requires the number of impurities to change. For example,
for the open-closed transition $H_{o\leftrightarrow c}$, ``almost''
energy conservation requires $n_{o}=n_{c}-1$, where $n_{r}$ is the
number of impurities of the $r$-th string. Likewise, ``almost''
energy conservation for the cubic open transition $H_{o\leftrightarrow
oo}$ requires $n_{oo}= n_{o}-1$.
We now turn to string
interactions and their gauge theory realization.

\newsec{Cubic Open String Field Theory}

The interaction Hamiltonian $H_{o\leftrightarrow oo}$ describing the
transition amplitude between a single string and a two-string state on
the D7-O7 system described in section $2$
can be obtained by generalizing the work in \GreenFU\ to the plane
wave background. This analysis has been recently carried out by
in  \ChandrasekharFQ\StefanskiZC . Here we evaluate the Hamiltonian for 
purely
bosonic string states. As
we show in Appendix F, the interaction vertex for this class of states
is given by\foot{We take without loss of generality
$\alpha^\prime p^+_{(3)}=-1, \alpha^\prime p^+_{(1)}=y$ and
$\alpha^\prime p^+_{(2)}=1-y$, where
$0<y<1$. Therefore, $\lambda^\prime=1/\mu^2$. Just as in
\PearsonZS\GomisWI\ the overall normalization $C=-i{\sqrt{ y(1-y)}}$
  is fixed by comparing to
one field theory amplitude.}
\eqn\cubicopen{{1\over \mu}|H\ra=CP|V\ra.}
$P$ is the prefactor,
\eqn\prefactor{
P=\sum_{r=1}^3\sum_{n=0}^\infty \bar F_{n(r)}(y)a_{n(r)}^{z'\dagger},}
and the explicit formulas for $\bar F_{n(r)}(y)$ are summarized in
Appendix
C.
$|V\ra$ describes the geometrical gluing of strings
\ifig\Seminearest{Cubic open string interaction diagram.}
{\epsfxsize1.5in\epsfbox{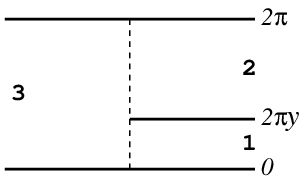}}
\noindent
and is given in terms
of a squeezed state\foot{Here and in section 5 
we omit an overall $p^+$ conservation
    factor, $|p^+_{(3)}|\ \delta(p^+_{(1)}+p^+_{(2)}+p^+_{(3)})$.}
\eqn\Neummm{
|V\ra=\exp\left({1\over 2}\sum_{r,s=1}^3\sum_{m,n}
a_{m(r)}^{I\dagger}\;\bar N_{m,n\;
IJ}^{(rs)}(y)\;a_{n(s)}^{J\dagger}\right)
|\vac\ra_{123},}
with $|\vac\ra_{123}=|\vac\ra_1\otimes|\vac\ra_2\otimes|\vac\ra_3$ and
where $|\vac\ra_r$
is the lowest energy open string state for the $r$-th
string. $\bar N_{m,n\;IJ}^{(r,s)}(y)$ are Neumann matrices which are
$SO(6)\times U(1)$ symmetric. The  non-vanishing entries are given by
$(m,n\geq 0)$:
\eqn\Neumannexpli{
\bar N_{m,n\; IJ}^{(rs)}(y)=\left\{\matrix{\delta_{I\bar{J}} \bar
N_{m,n}^{(rs)}(y)\qquad\quad I,J\in N,\cr
\noalign{\vskip15pt}
\delta_{I\bar{J}} \bar
N_{-m,-n}^{(rs)}(y)\qquad I,J\in D,}\right.}
and explicit expressions for $\bar N_{m,n}^{(rs)}(y),\bar
N_{-m,-n}^{(rs)}(y)$ in the large $\mu$ limit are summarized in
Appendix C and obtained by demanding worldsheet continuity in the
interaction diagram (Appendix B).

We note that the prefactor only involves a single bosonic oscillator
direction. This can be shown by properly imposing the
symmetries of the problem. The vacuum state $|\vac\ra$ is charged
\BerensteinZW\ under rotations in the ${\bf R}^2_{56}$ plane of
the transverse ${\bf R}^8$ geometry, in fact
\eqn\chargevac{
J^{56}|\vac\ra_{123}=-|\vac\ra_{123}.}
On the other hand the supersymmetry algebra requires that
\eqn\susyal{
J^{56}|H\ra=0,}
which selects a unique direction $I=z'$ in the prefactor
-- whose corresponding oscillator satisfies
$[J^{56},a^{z'\dagger}]=a^{z'\dagger}$ --
and therefore \susyal\ is realized.

We are now in the position of computing arbitrary Hamiltonian matrix
elements between single string and two-string states. Following
\GomisKJ\ it is convenient to introduce Feynman rules to evaluate
these amplitudes. They are given by\foot{Here the Neumann matrices
for the Neumann directions are different than those along the
Dirichlet directions, so we must keep track of the direction of the
oscillator. For later convenience we make explicit the fact that $
N^{(rs)}_{m,n\ IJ}(y)$ and $\bar F_{n(r)}(y)$ are functions of $y$.}:

\bigskip

\noindent
\eqn\Feynmm{\eqalign{
&(r,m,I)\;\hbox{---------------}\;(s,n,J)\qquad
\Longleftrightarrow\qquad \bar N^{(rs)}_{m,n\ IJ}(y),\cr
&(r,m,I)\;\hbox{---------------}
\!\!\times \qquad\qquad\ \;\;\;
\Longleftrightarrow\qquad \delta_{I,z^\prime}\bar F_{n(r)}(y),}}

\bigskip
\noindent
where $r,s\in\{1,2,3\}$ label the string, $m,n$ denote the worldsheet
momentum of the oscillator $a_{m(r)}^I$ and $I,J$ denote the direction
of the
oscillator.

The answer for the complete amplitude is obtained by gluing in all
inequivalent ways the single string state with the two-string
state.  There are two inequivalent gluings which differ in the ordering
of the strings:
\ifig\Seminearest{Contributions to amplitude.}
{\epsfxsize3.5in\epsfbox{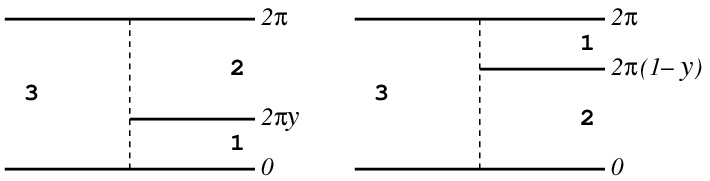}}
\noindent
The oscillator contribution to the first diagram can be
easily obtained using \Feynmm\
while the result for the second diagram can be easily obtained
from the result of the first diagram by making the replacements
$1\leftrightarrow
2$ and $y\leftrightarrow 1-y$. The interaction is non-vanishing
only if the
Chan-Paton index on the left end of one string is identical to the
Chan-Paton index on the right end of the neighboring
string. Therefore, the first diagram  in Fig. 2 comes with a factor of
    $\hbox{Tr}(\lambda^1\lambda^2\lambda^3)$ while the second diagram has
a factor of $\hbox{Tr}(\lambda^2\lambda^1\lambda^3)$.

We now compute the ``almost'' energy conserving amplitudes
described at the end of section $2$,  which we will reproduce from
perturbative gauge theory in the next section. These amplitudes are
between a single string state -- described by string 3 --
and a two-string state -- described by string 1 and 2 -- with the single
string state carrying one more impurity than the two-string state.
Therefore, we must evaluate the overlap of $|H\ra$ with $2n-1$
annihilation operators, where $n$ is the number of impurities in the
single string state. This overlap is computed by sequentially commuting
the
prefactor $P$ in \prefactor\ through each of the $2n-1$
oscillators. Given \Feynmm\ it follows that each of the $2n-1$ terms
is now multiplied by $\bar F_{(r)}$, where $r$ labels the string to which
the impurity was associated and which the prefactor annihilated. To
complete the calculation, we must
compute the matrix elements of the remaining $2n-2$ oscillators in
each of the $2n-1$ terms  with $|V\ra$. This is computed by contracting
in all
possible ways the $2n-2$ oscillators using the Neumann matrices as
propagators.

We can now classify which amplitudes are non-zero to leading order in
the large
$\mu$ expansion, that is to ${\cal O}(\lambda^\prime)$. Diagrams where
the prefactor acts on either string 1 or 2 are proportional to
$\bar F_{(r)} N^{(33)}$ where $r=1$ or $2$. Using the formulas in
Appendix C
we see that such diagrams are subleading in the large $\mu$
expansion. Therefore, to leading order, the prefactor must go though
an oscillator in string 3, and such diagrams are proportional to
$\bar F_{(3)}$. To leading order in the large $\mu$ expansion,
we must  contract the remaining $n-1$
oscillators in string 3 with the oscillators in string $1$ and
$2$. Diagrams with self-contractions, where two oscillators in string
3 are connected via $N^{(33)}$, are subleading. Moreover, since the
prefactor contains a bosonic creation operator only along the
$z^\prime$ direction, to get a non-zero answer we must have a $z^\prime$
impurity in string 3.
To summarize, the
leading order diagrams correspond to diagrams in which the prefactor
acts on string 3, there are no self-contractions  and at least one
impurity in string 3 is a $z^\prime$ oscillator.

The leading amplitudes with one or two
different impurities in string 3, which can be Neumann or Dirichlet, 
 are given by\foot{We recall that $\la
s_A|H|s_B\ra\equiv \la s_A|\otimes \la s_B|H\ra$.}

\noindent
$\bullet$ One Impurity:
\ifig\Seminearest{One Impurity Diagrams.}
{\epsfxsize3in\epsfbox{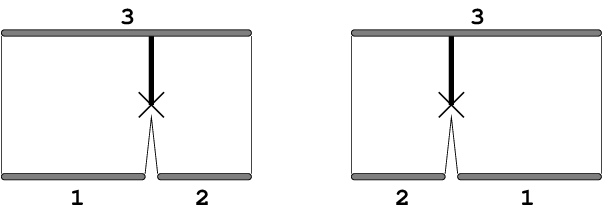}}
\eqn\explicit{\eqalign{
{1\over\mu}\la\la\vac,y|H|Z',n\ra
\simeq iC\left(\hbox{Tr}(\lambda^1\lambda^2\lambda^3)
\bar F_{n(3)}(y)+(1,y)\leftrightarrow (2,1-y)\right).}}

\noindent
$\bullet$ Additional Impurity in String 1:
\ifig\Seminearest{Diagrams for impurity in string 1.}
{\epsfxsize3in\epsfbox{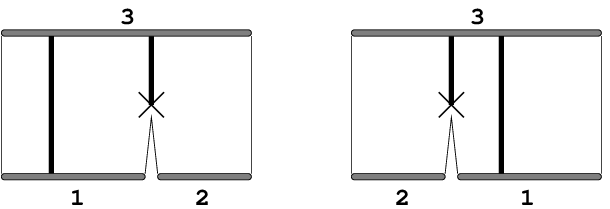}}
\eqn\explicita{\eqalign{
{1\over\mu}\la\la(Z',p)_1,y|
H|({\bar Z'},m\hskip-2pt:\hskip-2pt Z',n)\ra
&\hskip-2pt\simeq\hskip-2pt
-iC\hskip-2pt\left(\hskip-1pt\hbox{Tr}(\lambda^1\lambda^2\lambda^3)\bar 
F_{n(3)}(y)\bar
N^{(13)}_{p,m}(y)\hskip-2pt+\hskip-3pt(1,y)\leftrightarrow
(2,1\hskip-2pt-\hskip-2pt y)\hskip-1pt\right)\hskip-2pt,\cr
{1\over\mu}\la\la({\bar D},p)_1,y|
H|(D,m\hskip-2pt:\hskip-2pt Z',n)\ra
&\hskip-2pt\simeq\hskip-2pt
-iC\hskip-2pt\left(\hskip-1pt\hbox{Tr}(\lambda^1\lambda^2\lambda^3)\bar
F_{n(3)}(y)\bar
N^{(13)}_{-p,-m}(y)\hskip-2pt+\hskip-2pt
(1,y)\hskip-2pt\leftrightarrow\hskip-3pt
(2,1\hskip-2pt-\hskip-2pt y)\hskip-1pt\right)\hskip-2pt.}}

\noindent
$\bullet$ Additional Impurity in String 2:
\ifig\Seminearest{Diagrams for impurity in string 2.}
{\epsfxsize3in\epsfbox{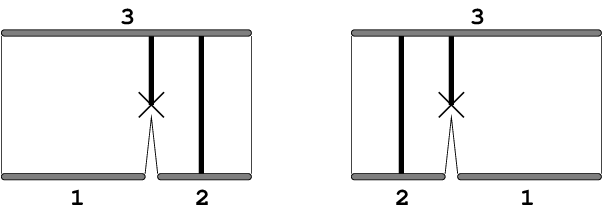}}
\noindent
\eqn\explicitb{\eqalign{
{1\over\mu}\la\la({Z'},p)_2,y|
H|({\bar Z'},m\hskip-2pt:\hskip-2pt Z',n)\ra
&\hskip-2pt\simeq\hskip-2pt
-iC\hskip-2pt\left(\hskip-1pt\hbox{Tr}
(\hskip-1pt\lambda^1\hskip-1pt\lambda^1\hskip-1pt\lambda^3\hskip-1pt)
\bar F_{n(3)}(y)\bar
N^{(23)}_{p,m}(y)\hskip-2pt+\hskip-2pt
(1,y)\hskip-2pt\leftrightarrow\hskip-2pt
(2,1-y)\hskip-1pt\right),\cr
{1\over\mu}\la\la({\bar D},p)_2,y|
H|(D,m\hskip-2pt:\hskip-2pt Z',n)\ra
&\hskip-2pt\simeq\hskip-2pt
-iC\hskip-2pt\left(\hskip-1pt\hbox{Tr}
(\hskip-1pt\lambda^1\hskip-1pt\lambda^2\hskip-1pt\lambda^3\hskip-1pt)
\bar F_{n(3)}(y)\bar
N^{(23)}_{-p,-m}(y)\hskip-2pt+\hskip-2pt
(1,y)\hskip-2pt\leftrightarrow\hskip-2pt
(2,1\hskip-2pt-\hskip-2pty)\hskip-1pt\right).}}
The numerical large $\mu$ expressions can be obtained by using the 
formulas in
Appendix C. We now derive these amplitudes using gauge theory.

\newsec{Gauge Theory Description of Cubic Open String Field Theory}

In this section, we perform an independent
     gauge theory analysis to reproduce
the previous cubic open string field theory results.
As explained in section 2, the open
     string Hamiltonian matrix elements
are to be captured by the ${\cal O}(\sqrt{g_2})$ contribution to the
matrix of anomalous dimensions of the BMN operators
\BMNopenvac\BMNopen\BMNopenb.
     Therefore, we must compute the
two-point functions of a single-string  BMN
     operator and a two-string BMN
     operator.

The term in the gauge theory Lagrangian responsible for the open string
interactions is given by:
\eqn\intera{
{\cal L}_{\rm int}=g^2 \bar Q^i\Omega[Z\Omega,
Z^\prime\Omega]\bar Q^i + c.c.\;.}
This interaction annihilates (or creates) two identical quarks and
creates (or annihilates) two bulk fields. From this coupling, we can
deduce the following selection rules.

\noindent
$\bullet$ {\it Selection Rule 1}:

The coupling in \intera\ is
$SO(8)$ invariant -- which is part of
     the flavor symmetry of the gauge
theory --, and so only the trace part of the $SO(8)$ indices of
the external quarks which interact via \intera\
in the BMN operators contribute.
Also, the quark propagator is
diagonal in $SO(8)$ indices.
Altogether, we obtain a  trace of
Chan-Paton wave functions along the
``boundary"\foot{By this,
we mean quark lines.} of the Feynman diagram.
This fits nicely
with the string theory expectation on
Chan-Paton wave functions.
     As summarized in the previous
section, open strings interact only if  the
Chan-Paton index on the left end of
     one string is identical to the
Chan-Paton index on the right end of the neighboring
string, thereby producing a trace.
     Furthermore, notice that there are
     two inequivalent ways of forming the Chan-Paton
     trace in the gauge theory computation
of two-point functions between
single-string and two-string BMN operators
depending on the order of two strings along the
single string. One gives
     $\hbox{Tr}(\lambda^1\lambda^2\lambda^3)$ and
the other $\hbox{Tr}(\lambda^1\lambda^3\lambda^2)$,
     where the $\lambda$'s are
the Chan-Paton wavefunctions
in \BMNopenvac\BMNopen\BMNopenb .
This nicely agrees with the previous string theory computation.

\noindent
$\bullet$ {\it Selection Rule 2}

As emphasized in section 2, the lowest order contribution to
     cubic open two-point function
     is the one-loop (${\cal O}(\lambda^\prime)$) amplitude. Since
     the quark number should change
     during the process, we should
insert one interaction vertex \intera\ to annihilate
     two quarks and create two bulk fields.
Therefore, one impurity in the single-string BMN operator must
be created by the annihilation of the quarks via \intera, and
it should be $Z^\prime$ because the interaction
term \intera\ can only make a $Z^\prime$ impurity.
Furthermore, if we limit ourselves to the leading order in
$\lambda^\prime$, all the rest of
     the impurities must freely contract.
This requires that each impurity
in the single-string BMN operator,
other than  $Z^\prime$ participating in \intera,
  should be paired with one of the impurities in the two-string
BMN operator in order to produce a non-vanishing amplitude.
Consequently,  all two-point functions
where the single-string BMN operator has
an unpaired Dirichlet impurity
or $\bar Z^\prime$ impurity vanish.
This selection rule nicely agrees with that of  string theory
shown in the previous section. Whenever the unpaired
single-open string impurity
is $\bar Z^\prime$ or along the Dirichlet directions,
      the cubic open string field theory Hamiltonian
matrix elements vanish for bosonic external states.
This is due to the fact that the prefactor in \prefactor\ only
contains the $Z^\prime$  impurity oscillator.

Another important consequence
of this consideration is that
there is no self-contraction contribution to leading order in
$\lambda^\prime$.
It was shown in \GomisKJ\ that the leading contribution of
     self-contractions in string field theory
     Feynman diagrams corresponds to
gauge theory interaction vertex involving four impurities. In our
case, we simply do not have room for this interaction vertex
after inserting \intera\ to leading order in $\lambda^\prime$.

\noindent
$\bullet$ {\it Explicit Computations}

Equipped with the two selection rules, let us now
perform explicit two-point function calculations.
     The factorization
     property of gauge theory Feynman diagrams
     in a dilute gas approximation
\GomisKJ\ suggests
     to consider each part of a Feynman diagram
     separately and put them together at the end.
Hence, we start by studying the simplest possible case
     and bring in more impurities to generalize the
analysis to arbitrary impurities following  \GomisKJ.

    From selection rule 2,
     the single-string BMN operator should
have a $Z^\prime$ impurity to split
     into a two-string BMN operator.
     The simplest amplitude is the
decay of a single open string
     with a $Z^\prime$ into two
vacuum open
strings.

$\bullet$ One Impurity:

\ifig\Seminearest{Cubic open gauge theory diagram.}
{\epsfxsize3in\epsfbox{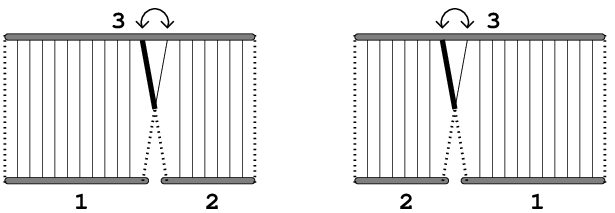}}

\eqn\otooovac{\eqalign{
&(4\pi^2|x|^2)^{J+2}\Big\la:O^{J_1}_{\vac}O^{J_2}_{\vac}:(x)\;
\overline{O^{J}}\!\!\!_{n,Z^\prime}(0) \Big\ra\cr
&=4{\sqrt{g_2}\over \sqrt{J}}\
\left\{  -{g^2 (2N) \over 8 \pi^2J}\sqrt{2}
\left[\cos\left({\pi n J_1\over J}\right)-
\cos\left({\pi n (J_1+1)\over J}\right)\right]{\rm Tr}
(\lambda^1\lambda^2\lambda^3) \right. \cr
&\quad \left.  -{g^2 (2N) \over 8 \pi^2J}\sqrt{2}
     \left[\cos\left({\pi n J_2 \over J}\right)-
\cos\left({\pi n (J_2+1)\over J}\right)\right]{\rm Tr}
(\lambda^2\lambda^1\lambda^3) \right\}
\ln(x^2\Lambda^2)^{-1},}}
where $J_1+J_2=J$ and $y=J_1/J$.
As mentioned above, we have two diagrams depending
on the order of open string operator 1 and 2 along operator 3,
each giving ${\rm Tr}(\lambda^1\lambda^2\lambda^3)$ and
${\rm Tr}(\lambda^2\lambda^1\lambda^3)$ respectively.
  Furthermore, given a diagram, we have two choices of
propagators to be used for
overlapping fields in string 1 and string 3 as well as for
overlapping fields in string 2 and string 3.
Using the twisted propagator for all fields in an operator
is equivalent to transposing the operator and using the
ordinary propagator instead. In addition,
open string BMN operators are invariant under transposition.
Therefore, all of the four
cases give the same result and we have a factor of $4$ in
\otooovac.

The phase difference inside $[\cdot]$
comes from the commutator of
interaction \intera\  and the rest is the result of the
loop diagram, where in particular:
\eqn\loopint{
\left({1\over 4\pi^2}\right)^4\int d^4y{1\over y^4 (x-y)^4}=-{1\over
8\pi^2}\left({1\over 4\pi^2|x|^2}\right)^2\ln(x^2\Lambda^2)^{-1}.}
We can now relate in the BMN limit the phase difference  to the
large $\mu$ behaviour of the string field theory prefactor:
\eqn\Fplusthree{ \eqalign{
    -{g^2 (2N) \over 2\pi^2J}\sqrt{2}
\left[\cos\left({\pi n J_1\over J}\right)-
\cos\left({\pi n (J_1+1)\over J}\right)\right]
    &\simeq  \bar F_{n(3)}(y),\cr
    -{g^2 (2N) \over 2\pi^2J}\sqrt{2}
\left[\cos\left({\pi n J_2\over J}\right)-
\cos\left({\pi n (J_2+1)\over J}\right)\right]
    &\simeq  \bar F_{n(3)}(1-y),
}}
which reproduces the string field theory  prefactor in Appendix C.
After taking into account the factor 
   $\sqrt{Jy(1-y)}$, explained in section 2, that
we need to go from gauge theory to string theory\foot{From now on, 
whenever we
write $\Gamma$ in this section this factor will have already been taken
into account.},
   the anomalous dimension matrix is expressed as
\eqn\Gammaotooovac{
\Gamma^{(1/2)}_{(n,Z^\prime)\lambda^3,
\;{\rm vac};y\lambda^1\lambda^2} \simeq
{\sqrt{y(1-y)}} \left[ \bar F_{n(3)}(y)
{\rm Tr}(\lambda^1\lambda^2\lambda^3)
    +(1,y)\leftrightarrow(2,1-y)\right],}
which reproduces the string theory computation \explicit.

Now let us consider a process with one
more impurity along string 1 and string 2 respectively. The impurity
can be either along a
Neumann or Dirichlet direction.

$\bullet$ Additional Impurity in String 1:
\ifig\Seminearest{Diagrams with extra impurity in string 1.}
{\epsfxsize3in\epsfbox{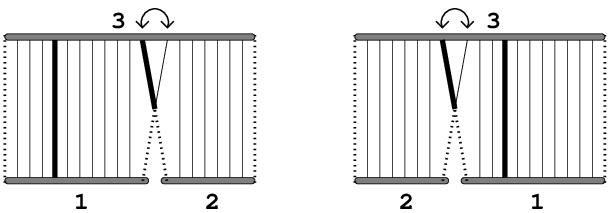}}

If we add a Neumann impurity with momentum $p$ and $m$ respectively
along string 1 and 3, we need to multiply the loop factor in
\otooovac\ by the sum over free contractions of the new impurity. As
before, there are two possibilities depending on the ordering 
 of the strings. In
the BMN limit we can reexpress the sum over phases for the two
orderings as the integral
representation of the string field theory Neumann matrices(see
Appendix C):
\eqn\freeNone{\eqalign{
    {2\over \sqrt{JJ_1}} \sum_{l=0}^{J_1-1}
\cos\left({\pi p l \over J_1}\right)
     \cos\left({\pi m l \over J}\right)
\simeq  -\bar N^{(13)}_{p,m}(y)
& \qquad \hbox{for }\;
    {\rm Tr}(\lambda^1\lambda^2\lambda^{3}), \cr
    {2\over \sqrt{JJ_1}} \sum_{l=0}^{J_1-1}
\cos\left({\pi p l \over J_1}\right)
     \cos\left({\pi m (J_2+l) \over J}\right)
\simeq  -\bar N^{(23)}_{p,m}(1-y)
& \qquad \hbox{for }\;
{\rm Tr}(\lambda^2\lambda^1\lambda^{3}).
}}
Therefore, the final result is
\eqn\otoooZp{\eqalign{
&\Gamma^{(1/2)}_{(n,Z^\prime)(m,\bar Z^\prime)\lambda^3
,\;(p,\bar Z^\prime)y\lambda^1\lambda^2}\cr
&\simeq-{\sqrt{y(1-y)}}
    \left[ {\rm Tr}(\lambda^1\lambda^2\lambda^3)\bar F_{n(3)}(y)
\bar N^{(13)}_{p,m}(y)
+(1,y)\leftrightarrow(2,1-y) \right],
}}
which matches the string theory expectation \explicita.

If the extra impurity is in the  Dirichlet direction,  we need to
multiply the loop factor in
\otooovac\ by the sum over free contractions of the new impurity for
the two orderings,
which yields in the BMN limit the integral representation of the
Neumann matrices:
\eqn\freeDone{\eqalign{
    {2\over \sqrt{JJ_1}} \sum_{l=0}^{J_1-1}
\sin\left({\pi p l \over J_1}\right)
     \sin\left({\pi m l \over J}\right)
\simeq  -\bar N^{(13)}_{-p,-m}(y)
& \qquad \hbox{for }\;
{\rm Tr}(\lambda^1\lambda^2\lambda^{3}), \cr
    {2\over \sqrt{JJ_1}} \sum_{l=0}^{J_1-1}
\sin\left({\pi p l \over J_1}\right)
     \sin\left({\pi m (J_2+l) \over J}\right)
\simeq -\bar N^{(23)}_{-p,-m}(1-y)
& \qquad \hbox{for }\;
    {\rm Tr}(\lambda^2\lambda^1\lambda^{3}).
}}
The final result is
\eqn\otoooD{\eqalign{
&\Gamma^{(1/2)}_{(n,Z^\prime)(m,D)\lambda^3,
\;(p,D)y\lambda^1\lambda^2}\cr
&\simeq -{\sqrt{y(1-y)}}
    \left[ {\rm Tr}(\lambda^1\lambda^2\lambda^3)\bar F_{n(3)}(y)
\bar N^{(13)}_{-p,-m}(y)
+(1,y)\leftrightarrow(2,1-y) \right],
}}
and exactly reproduces \explicita.

\vfill\eject

$\bullet$ Additional Impurity in String 2:

\ifig\Seminearest{Diagrams with extra impurity in string 2.}
{\epsfxsize3in\epsfbox{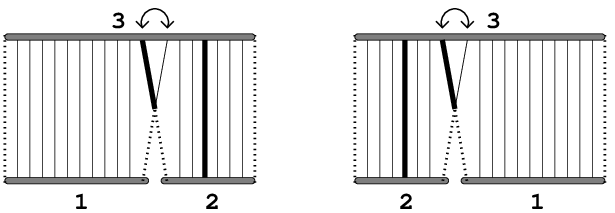}}

If we add a Neumann impurity with momentum $p$ and $m$ respectively
along string 2 and 3, we need to multiply the loop factor in
\otooovac\ by the sum over free contractions of the new impurity:
\eqn\freeNone{\eqalign{
    {2\over \sqrt{JJ_2}} \sum_{l=0}^{J_2-1}
\cos\left({\pi p l \over J_2}\right)
     \cos\left({\pi m (l+J_1) \over J}\right)
\simeq  -\bar N^{(23)}_{p,m}(y)
& \qquad \hbox{for }\;
    {\rm Tr}(\lambda^1\lambda^2\lambda^{3}), \cr
    {2\over \sqrt{JJ_2}} \sum_{l=0}^{J_2-1}
\cos\left({\pi p l \over J_2}\right)
     \cos\left({\pi m l \over J}\right)
\simeq  -\bar N^{(13)}_{p,m}(1-y)
& \qquad \hbox{for }\;
{\rm Tr}(\lambda^2\lambda^1\lambda^{3}).
}}
Therefore, the final result  result is
\eqn\otoooZp{\eqalign{
&\Gamma^{(1/2)}_{(n,Z^\prime)(m,\bar Z^\prime)\lambda^3
,\;(p,\bar Z^\prime)y\lambda^1\lambda^2}\cr
&\simeq-{\sqrt{y(1-y)}}
    \left[ {\rm Tr}(\lambda^1\lambda^2\lambda^3)\bar F_{n(3)}(y)
\bar N^{(23)}_{p,m}(y)
+(1,y)\leftrightarrow(2,1-y) \right],
}}
which matches the string theory expectation \explicitb.

If the extra impurity is in the  Dirichlet direction,  we need to
multiply the loop factor in
\otooovac\ by the sum over free contractions of the new impurity:
\eqn\freeDone{\eqalign{
    {2\over \sqrt{JJ_2}} \sum_{l=0}^{J_2-1}
\sin\left({\pi p l \over J_2}\right)
     \sin\left({\pi m (l+J_1) \over J}\right)
\simeq  -\bar N^{(23)}_{-p,-m}(y)
& \qquad \hbox{for }\;
{\rm Tr}(\lambda^1\lambda^2\lambda^{3}), \cr
    {2\over \sqrt{JJ_2}} \sum_{l=0}^{J_2-1}
\sin\left({\pi p l \over J_2}\right)
     \sin\left({\pi m l \over J}\right)
\simeq -\bar N^{(13)}_{-p,-m}(1-y)
& \qquad \hbox{for }\;
    {\rm Tr}(\lambda^2\lambda^1\lambda^{3}).
}}
The computation is then straightforward:
\eqn\otoooD{\eqalign{
&\Gamma^{(1/2)}_{(n,Z^\prime)(m,D)\lambda^3,
\;(p,D)y\lambda^1\lambda^2}\cr
&\simeq -{\sqrt{y(1-y)}}
    \left[ {\rm Tr}(\lambda^1\lambda^2\lambda^3)\bar F_{n(3)}(y)
\bar N^{(13)}_{-p,-m}(y)
+(1,y)\leftrightarrow(2,1-y) \right],
}}
and exactly reproduces \explicitb.

The generalization to arbitrary impurities is now straightforward along
the lines of \GomisKJ. One of the
$Z^\prime$ impurities in the single-string BMN operator
must participate
    in the interaction vertex \intera\ yielding
$\bar F_{n(3)}(y)$ and
the rest of impurities should be
    paired up between the single-string
    and the two-string BMN operators.
    Then, each pair of Neumann
impurities results in  a factor of
$\bar N^{(r3)}_{p,m}(y)$ while each pair of
Dirichlet impurities produces a factor of $\bar N^{(r3)}_{-p,-m}$,
where
$r=1$ or $2$ depending on the location of the impurity in the
two-string BMN operator. This is the result for the diagram
proportional to  $\hbox{Tr}(\lambda^1\lambda^2\lambda^3)$. The result
of the diagram proportional to   $\hbox{Tr}(\lambda^1\lambda^3\lambda^2)$
  can be obtained by the replacement $(1,y)\leftrightarrow(2,1-y)$ and
this exactly agrees with the string theory result.
Both self-contractions in string theory  and its gauge theory counterpart
do not appear  to leading order in $\lambda^\prime$,
    making the discussion simpler than in \GomisKJ, where they played
an important role.
Altogether, the gauge theory answer reproduces the result
of the corresponding string field theory Feynman diagram.

\newsec{Open-Closed String Field Theory}

In this section we construct the interaction
Hamiltonian $H_{o\leftrightarrow c}$ describing the amplitude of an
open string in the D7-O7 system to annihilate into a closed
string. This process occurs when the ends of the open string join to
make a closed string. The construction of $H_{o\leftrightarrow c}$ can
be obtained following the work in \GreenFU . The overlap part of the
vertex
is by demanding continuity of all worldsheet fields on the
interaction diagram:
\ifig\Seminearest{Open-Closed Interaction Diagram.}
{\epsfxsize2in\epsfbox{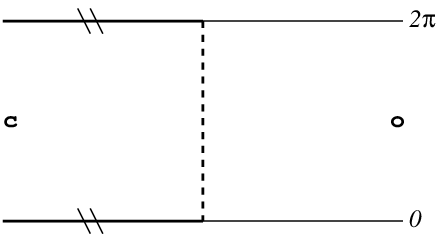}}

   From the gluing of worldsheet fields\foot{We have taken without loss
of generality $\alpha^\prime p^{+}_{c}=-\alpha^\prime p^{+}_{o}=1$.}
we can construct the squeezed
state $|V\ra$. The
construction of the Neumann matrices and the proof of certain
factorization theorems are relegated to the Appendices D,G. The final
ingredient in finding  $H_{o\leftrightarrow c}$ is to find the
appropriate prefactor. The prefactor must both preserve all the
kinematical symmetries\foot{The kinematical symmetries are the
symmetries that preserve the light-cone and kappa symmetry gauge
choice while
the dynamical ones do not preserve it and require compensating gauge
transformations to leave the gauge choice invariant.} and implement
the superalgebra for the
dynamical symmetries.

As explained in section $2$, the local interaction for
$H_{o\leftrightarrow c}$ is identical to that in $H_{o\leftrightarrow
oo}$. This suggests that the form of the prefactor for these
interactions are identical, an expectation that was beautifully proven
in \GreenFU\ for flat space. 
Moreover, the recent analysis of
$H_{o\leftrightarrow oo}$
in \StefanskiZC\ has moreover confirmed that the form of
the prefactor
in the plane wave background is identical to that in flat
space. Physically, the origin of the
equivalence is that the details of the prefactor depend only on the
short distance properties on the worldsheet.
Therefore, we will use the general form of the prefactor for
the ``joining-splitting'' type of interaction which can be found in
\GreenFU\StefanskiZC . It would be interesting to confirm this
physical input by performing the analysis of the supersymmetry algebra
for the open-closed vertex.

We are now in a position to evaluate the Hamiltonian for arbitrary
bosonic string states. The Hamiltonian is given by\foot{As section 3,
the overall normalization $C=-i/\sqrt{2}$ is fixed by comparing with one
gauge theory result.}
\eqn\opcl{{1\over\mu}|H\ra= CP|V\ra,}
where $P$ is the prefactor
\eqn\prefactoc{
P=\sum_{r=o,c}\sum_{n=0}^\infty
F_{n(r)}a^{z^\prime\dagger}_{n(r)},}
and the formula  for $F_{n(r)}$ is given in Appendix D. Note that the
prefactor is structurally the same as in $H_{o\leftrightarrow oo}$.
$|V\ra$ describes the geometrical gluing of strings computed in
Appendix B
\eqn\overlapoc{
|V\ra=\exp\biggl({1\over 2}\sum_{r,s=o,c}\sum_{m,n}
a^{I\dagger}_{m(r)}N^{(rs)}_{mn\ IJ}a^{J\dagger}_{n(r)}\biggr)
|\vac\ra_{oc},}
with $|\hbox{vac}\ra_{oc}=|\hbox{vac}\ra_{o}\otimes|\hbox{vac}\ra_{c}$
and where  $|\hbox{vac}\ra_{r}$ is the lowest energy state of the
$r$-th string. For this vertex,  the closed string oscillator
$a_{n(c)}^I\equiv \alpha_{n}^I$
index $n$ runs
over all integers, while
for the open string oscillator $a_{n(o)}^I$ the index $n$ runs over only
non-negative ($\cos$) modes for the Neumann directions while it runs
over negative ($\sin$) modes for Dirichlet directions.
    $N^{(rs)}_{mn\ IJ}$ are therefore $SO(6)\times U(1)$
symmetric and are described in Appendices B,D. We would like to note that
the $U(1)_{56}$ symmetry of the Hamiltonian isolates the $z^\prime$
direction in the prefactor, the proof of which is identical to that
presented in
section $3$ for $H_{o\leftrightarrow oo}$.

Simple Feynman rules can be extracted by the Hamiltonian, they are
given by
\bigskip

\noindent
\eqn\Feynmmoc{\eqalign{
&(r,m,I)\;\hbox{---------------}\;(s,n,J)\qquad
\Longleftrightarrow\qquad  N^{(rs)}_{m,n\ IJ},\cr
&(r,m,I)\;\hbox{---------------}
\!\!\times \qquad\qquad\ \;\;\;
\Longleftrightarrow\qquad \delta_{I,z^\prime}F_{m(r)}.}}

\bigskip
\noindent
Using the large $\mu$ formulas for $N^{(rs)}_{mn\ IJ}$ and
$F_{n(r)}$ in Appendix D and the logic in section $3$ we
    can also show that to leading order in the
$\lambda^\prime$ expansion -- that is to ${\cal O}(\lambda^\prime)$ --
that the diagrams that contribute are those in which the prefactor
acts on a closed string oscillator, there are no self-contractions and
at least one of the impurities in the closed string is a $z^\prime$
oscillator.

Any amplitude has two basic ingredients. One comes from contracting
all oscillators using the Feynman rules in \Feynmmoc . The other
contribution summarizes the well known result that the open string can
transform into a closed string if the two ends of the open string carry
the
same Chan-Paton factor indices. The amplitude is therefore multiplied by
$\hbox{Tr}(\lambda)$, where $\lambda$ is the Chan-Paton wavefunction
of the open string state. Since we are working with unoriented
strings, whose Chan-Paton factors satisfy the symmetry properties in
\parity\CPsymm,  it follows that amplitudes in which the total worldsheet
momentum of the open string is even vanish (since  then
$\lambda^\T=-\lambda$) while when the total open string worldsheet
momentum is odd only the trace part  of the $\lambda$ matrix
contributes (since then
$\lambda^\T=\lambda$).

The leading amplitudes are given by\foot{Here we relax the level
matching condition since then it is easier to generalize to
higher impurities. If one wants only two impurities one just sets 
$p=-m$.}

\noindent
$\bullet$ Neumann Contraction:
\eqn\example{\eqalign{
{1\over\mu}\la (Z',m:{\bar N},p)|H|N,n\ra
&\simeq -i\sqrt{2}C\; \hbox{Tr}(\lambda) F_{m(c)}N^{(co)}_{p,n},}}
where we have used the symmetry properties of $F_{m(c)}$ and
$N^{(co)}_{p,n}$ for the contribution due to the second term in
\BMNclo\ required by $\Omega$-invariance.

\noindent
$\bullet$ Dirichlet Contraction:
\eqn\examplea{\eqalign{
{1\over\mu}\la (Z',m:{\bar D},p)|H|D,n\ra
&\simeq -i\sqrt{2}C\; \hbox{Tr}(\lambda) F_{m(c)}N^{(co)}_{p,-n}},}
where we have rewritten the second contribution using the formulas in
Appendix D.

The diagram for both processes look the same and is given by:

\ifig\Seminearest{Open-Closed Interaction Diagram.}
{\epsfxsize2in\epsfbox{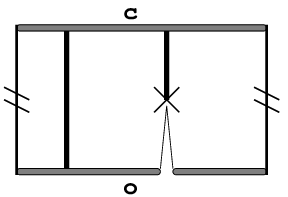}}

Adding more impurities -- in an ``almost'' energy conserving way --
multiplies \example\examplea\  by the corresponding
Neumann matrices. We now turn to the
gauge theory derivation.

\newsec{Gauge Theory Description of Open-Closed String Field Theory}

In this section, we analyze the two-point functions
    of open and closed string BMN
operators to reproduce the open-closed
string field theory result in the previous
section.  The two selection rules in Section 4
also are very useful in this computation.
From selection rule 1, we conclude that the amplitude is proportional
to ${\rm Tr}(\lambda)$ where
$\lambda$ is the Chan-Paton wave function
    of the open string BMN operator. Therefore,
    the sum of worldsheet momenta of impurities
    in the open string BMN operator should be
odd \CPsymm\ in order to have a non-vanishing  amplitude.
    Also from selection rule 2,
    the closed string BMN operator should have a
$Z^\prime$ impurity to interact with the open string BMN operator. One
$Z^\prime$ impurity is annihilated with one $Z$ to create two
quarks via \intera.
Moreover, the rest of the impurities in the closed string BMN operator
must be paired up with impurities in the open string BMN operator and
freely contract to leading order on $\lambda^\prime$.

\noindent
$\bullet$ Neumann Impurity:

First, let us consider the simple case that the closed string BMN operator
    has $Z^\prime$ and a Neumann impurity $N$. Accordingly,
    the open string BMN operator has the same impurity $N$. The amplitude 
is given by\foot{As in the previous section, we relax the level matching
condition, we can instate by letting $p=-m$.} 
\eqn\otocD{ \eqalign{
&(4\pi^2|x|^2)^{J+2} \la O^{J}_{n,N}(x)\;
\bar O^J_{(Z^\prime,m:N,p)}(0) \ra\cr
&=2\sqrt{g_2}\left[ {\sqrt{2}\over J} \sum_{l=0}^{J-1} e^{{-2\pi i p
l\over J}}
     \cos\left({\pi n l \over J}\right) \right]
\left[{g^2(2N)\over 8 \pi^2J}\left( e^{-{2\pi i m \over
J}}-1\right)
\right]
{\rm Tr}(\lambda) \ln(x^2 \Lambda^2)^{-1},}}
where the first factor comes from the free contraction of the $N$
impurity and the rest from the interaction vertex \intera.
Here, we also have  two choices of propagators -- twisted
and untwisted -- in the planar limit. As explained in section 4,
when we use the twisted propagator for all fields, we can instead
transpose the open string BMN operator and use the ordinary
propagator for all fields. Then, we have the same result as obtained
  with the ordinary propagator because open string BMN operators
  are invariant under transposition. Hence, we have a factor of 2 in
  \otocD.

As in Section
4,
we can identify each factor with objects in the
open-closed string field theory summarized in Appendix D.
The effect of the interaction term \intera\ is:
\eqn\preopclo{
{g^2(2N)\over 4 \pi^2J}\left( e^{-{2\pi i m \over
J}}-1\right)\simeq F_{m(c)}.}
The sum over free contractions yields the large $\mu$ Neumann matrix
\eqn\idenoc{\eqalign{
\left[ {\sqrt{2}\over J} \sum_{l=0}^{J-1} e^{{-2\pi i p l\over J}}
     \cos\left({\pi n l \over J}\right) \right]  &\simeq - N^{(co)}_
{p,n}.}}
Therefore,
\eqn\GammaotocD{
\Gamma^{(1/2)}_{(m,Z^\prime)(p,N),\;(n,N) \lambda}
\simeq -F_{m(c)}N^{(co)}_{p,n}
    {\rm Tr}(\lambda),}
which reproduces the string theory answer \example.

\ifig\Seminearest{Open-Closed Gauge Thery diagram.}
{\epsfxsize2in\epsfbox{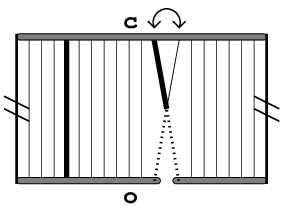}}

\noindent
$\bullet$ Dirichlet Impurity:

Let us now consider the simple case that the closed string BMN operator
    has $Z^\prime$ and a Dirichlet impurity $D$. Accordingly,
    the open string BMN operator has the same impurity $D$. The amplitude
is given by:
\eqn\otocD{ \eqalign{
&(4\pi^2|x|^2)^{J+2} \la O^{J}_{n,D}(x)\;
\bar O^J_{(Z^\prime,m:D,p)}(0) \ra\cr
&=\sqrt{g_2}\left[ {\sqrt{2}\over J} \sum_{l=0}^{J-1} e^{{-2\pi i p
l\over J}}
     \sin\left({\pi n l \over J}\right) \right]
\left[{g^2(2N)\over 4 \pi^2J}\left( e^{-{2\pi i m \over
J}}-1\right)
\right]
{\rm Tr}(\lambda) \ln(x^2 \Lambda^2)^{-1},}}
where the first factor comes from the free contraction of the $D$
impurity and the rest from the interaction vertex \intera.
The interaction terms is just as before while the sum over free
contraction now yields
\eqn\idenoc{\eqalign{
\left[ {\sqrt{2}\over J} \sum_{l=0}^{J-1} e^{{-2\pi i p l\over J}}
     \sin\left({\pi n l \over J}\right) \right]  &\simeq - N^{(co)}_
{p,-n}.}}
Therefore,
\eqn\GammaotocD{
\Gamma^{(1/2)}_{(m,Z^\prime)(p,D),\;(n,D) \lambda}
\simeq -F_{m(c)}N^{(co)}_{p,-n}
    {\rm Tr}(\lambda),}
which reproduces the string theory answer \examplea.

The generalization to arbitrary impurities is simpler than
   the cubic open interaction case. For each extra pair of impurities,
   we must add
   a Neumann matrix to the string answer. In gauge theory we must also
   sum over the new set of free contractions, but we have already proven
   that those yield the corresponding large $\mu$ Neumann
   matrices. Therefore, gauge theory reproduces string theory diagram
   by diagram.

\newsec{Conclusions}
In this paper we have analyzed the duality between  open+closed
string theory in the plane wave background and ${\cal N}=2$ $Sp(N)$
gauge theory with hypermultiplets in the $\anti\oplus 4\fund$
representations of $Sp(N)$. Combined with the results in \GomisKJ,
we have found that the proposal
\identifiint\ holds for
both ``joining-splitting'' type interactions and ``exchange''
interactions.  In establishing the correspondence, we have used the
conjectured mapping between string and gauge theory states proposed in
\GomisWI, which has proven universal in that it applies to all
interactions and is found without invoking string theory.

We have found that the individual ingredients of string field theory,
namely the prefactor and the Neumann matrices  can be separately
written in purely gauge theory terms. The ``exchange'' type
interaction prefactor is captured \GomisKJ\ by the action of a quartic 
F-term interaction  in gauge theory while the ``joining-splitting'' type
of interaction is captured by a quartic F-term interaction involving
quarks:
\eqn\mapinte{\eqalign{
{\cal L}_{\rm int}=
g^2 \bar Q^i\Omega[Z\Omega,Z^\prime\Omega]
\bar Q^i &\ \leftrightarrow
\ \hbox{``joining-splitting'' interaction},\cr
{\cal L}_{\rm int}=-g^2{\rm Tr}\left(
[Z\Omega,Z^\prime\Omega][\bar Z\Omega,
  \bar Z^\prime\Omega]\right)
&\ \leftrightarrow
\ \hbox{``exchange'' interaction}.}}
  This has been shown 
explicitly for $H_{c\leftrightarrow cc},
H_{o\leftrightarrow oo}$ and $H_{o\leftrightarrow c}$. It is
straightforward to show that the prefactor for the rest of the
``exchange'' type 
interactions are captured by  the aforementioned gauge theory
interaction. For example, the rearrangement interactions  
$H_{c\leftrightarrow c}$ and $H_{o\leftrightarrow o}$ -- only existing
in unoriented string theory -- are computed by
gauge theory Feynman diagrams with ${\bf RP}^2$ topology with the
insertion of the F-term. The interaction reproduces the prefactor
while the Neumann matrices are computed by the gauge theory free contractions.
 Moreover, in Appendix B we have shown that
for all interactions that the large $\mu$ Neumann matrices are
precisely reproduced by the sum over free contractions in the gauge
theory:
\eqn\sumphase{
\hbox{sum over free contractions for one impurity}\leftrightarrow
\ \hbox{Neumann matrix}.}
The equivalence works diagram by diagram.

We should mention that other proposals have appeared in the literature
for computing matrix elements of the string Hamiltonian. These
proposals, formulated only to leading order in $\lambda^\prime$ and for
$H_{c\leftrightarrow cc}$, 
identify the string Hamiltonian matrix elements with the three point functions 
in gauge theory. We think, however, that the reported agreement is due to 
the use of an invalid string field theory Hamiltonian. The string Hamiltonian 
should respect all the symmetries of the plane wave background and should 
yield in the $\mu \rightarrow 0$ limit the correct flat space string theory 
vertex   
in \GreenHW, which reproduces the flat space amplitudes computed
using CFT techniques. 
Symmetries alone do not fix the vertex to leading order
and we must use the extra requirement 
that the correct flat space vertex is reproduced. In particular the vertex 
constructed in \DiVecchiaYP\ does not yield the correct flat space vertex in 
\GreenHW\ and does not preserve the $Z_2$ symmetry of the plane wave
 background, as recently emphasized by \PankiewiczKJ. Moreover, 
the identification of string theory and gauge theory 
Hamiltonians \identifiint\ makes it clear that the 
string theory Hamiltonian matrix elements are computed 
from two-point functions and not three-point functions. Gauge theory 
three-point functions are more naturally associated\foot{We would like 
to thank Juan Maldacena for discussions on this point.}
 with matrix elements between in and out states of a string vertex
operator (see also \MannQP).

Despite the  successful computations carried in this paper, 
it is not yet known how to  transcribe  the BMN
sector of gauge theory in terms of a complete theory, without any
truncation. Finding this theory is an important and challenging problem
that must be resolved in order to understand the underpinnings of holography 
in the plane wave background.

\bigbreak\bigskip\bigskip\centerline{{\bf Acknowledgements}}\nobreak
We would like to thank Hiroyuki Hata, Ken Intriligator, Juan Maldacena,
Hirosi Ooguri, John Schwarz
and Cumrun Vafa for useful
discussions.
J.~G.~is supported by a Sherman Fairchild Prize Fellowship. S.~M.~is
supported in part by JSPS Postdoctoral Fellowships for
Research Abroad H14-472.
This research was supported in part by the DOE grant
DE-FG03-92-ER40701.

\appendix{A}{Convention for $D=4\ {\cal N}=2\ Sp(N)$ Gauge Theory}
In this appendix, we fix our convention for $D=4\ {\cal N}=2\ Sp(N)$
     gauge theory. In the following, proper raising and lowering of indices with the invariant $Sp(N)$ tensor $\Omega_{ab}$ is assumed. 

\noindent
$\bullet$ {\it Lagrangian in Euclidean signature}
\eqn\Lag{\eqalign{
{\cal L} =& {1\over 2g^2} \hbox{Tr}\left( {1\over 4}
F_{\mu\nu}F^{\mu\nu}
+ \overline{D_\mu W} D^\mu W+ \overline{D_\mu Z}D^\mu Z
+ \overline{D_\mu Z^\prime}D^\mu Z^\prime\right)\cr
&+ {1\over g^2}  \overline{D_\mu Q^i} D^\mu Q^i\cr
&+{1\over 4g^2} \hbox{Tr}\left( [ \overline{W},W]+[ \overline{Z},Z]
+[ \overline{Z}^\prime,Z^\prime] - 2 Q^i\cdot  \overline{Q}^{i}\right)^2
\cr
&-{1\over g^2} \hbox{Tr}\left( [Z,Z^\prime][ \overline{Z},
\overline{Z}^\prime]
     + [W,Z][ \overline{W}, \overline{Z}]
+  [W,Z^\prime][ \overline{W}, \overline{Z}^\prime] \right) \cr
&+{1\over g^2}\left(  \overline{Q}^i[Z,Z^\prime] \overline{Q}^i
     + Q^i[\overline{Z}^\prime,\overline{Z}]Q^i +(\overline{Q}^iQ^i)^2
\right)
     +{2\over g^2} \overline{Q}^i \overline{W} W Q^i }}

\noindent
$\bullet$ {\it Propagators}
\eqn\propa{\eqalign{
\la {W_a}^b(x)\; {\bar W}_c^{\;\;d}(0) \ra &=
{g^2 \over 4\pi^2 |x|^2} \left(\delta_a^d\delta_c^b
     +\Omega_{ac}\Omega^{bd}\right) \cr
\la {Z_a}^b(x)\; {\bar Z}_c^{\;\;d}(0) \ra=\la {Z^{\prime\;b}_a}(x)\;
{\bar Z^{\prime\;d}_c}(0) \ra
     &= {g^2 \over 4\pi^2 |x|^2} \left(\delta_a^d\delta_c^b
     -\Omega_{ac}\Omega^{bd}\right) \cr
\la Q^i_a(x)\; {\bar Q}^{j b}(0)\ra &= {g^2 \over 4\pi^2 |x|^2}
\delta^{ij}\delta_a^b
}}

\appendix{B}{Large $\mu$ Neumann Matrices as Gauge Theory Free 
Contractions}

In this Appendix we show that the bosonic Neumann matrices for each of
the seven
interaction terms of
open+closed string field theory reduce in the large $\mu$ limit to the
Fourier overlaps relating the {\it out} state(s) oscillators to the
{\it in} state(s) oscillators on the corresponding Mandelstam
diagram. One can also show that the ``diagonal'' Neumann matrices
relating {\it in}({\it out}) state oscillators to {\it in}({\it
out}) state oscillators are subleading in the large $\mu$ limit.

The Neumann matrices are obtained by demanding continuity of the
worldsheet fields along the interaction diagram\foot{The interaction
$H_{oo\leftrightarrow oo}$ has two outgoing strings, but the argument
we present can be easily generalized to that interaction.}:
\eqn\conti{\eqalign{
\left(X_{1}(\sigma_{1})+X_2(\sigma_2)-X_3(\sigma_3)\right)|V\ra=0,\cr
\left(P_1(\sigma_1)+P_2(\sigma_2)+P_{3}(\sigma_{3})\right)|V\ra=0,}}
where the index $3$ denotes the {\it outgoing} string  and
${1,2}$ the {\it incoming} strings (for interactions where there is a
single incoming string we just set $X_2=P_{2}=0$). $\sigma_{i}$
describes the parametrization of the Mandelstam diagram. In order to
compute the Neumann matrices we need the worldsheet expansion of open
and closed strings, which we take with $\qquad\qquad0\leq \sigma\leq 2\pi \alpha$, 
where  $\alpha=\alpha^\prime p^+$, so that the mode frequencies are
$\w_n=\sqrt{(\mu\alpha)^2+n^2}$ and
$\bw_n=\sqrt{(\mu\alpha)^2+{n^2\over 4}}$. They are given by: 

\noindent
$\bullet\ $Closed String:

\eqn\closedexp{\eqalign{
x(\sigma)&=
\sum_{n=-\infty}^\infty i\sqrt{{\alpha^\prime\over 2\w_n}}
\left[\alpha_n-\alpha^\dagger_{-n}\right] e^{in{\sigma\over 
\alpha}},\cr
p(\sigma)&={1\over 2\pi\alpha}
\sum_{n=-\infty}^\infty
\sqrt{{\w_n\over 2\alpha^\prime}}
\left[\alpha_n+\alpha^\dagger_{-n}\right] e^{in{\sigma\over
\alpha}}.}}

\noindent
$\bullet\ $Open String:

\noindent
1)Neumann Boundary Condition:
\eqn\openexp{\eqalign{
x(\sigma)&=i\sqrt{{\alpha^\prime\over 2{\bw}_0}}(a_0-a_0^\dagger)+
\sum_{n=1}^\infty i\sqrt{{\alpha^\prime\over{\bw}_n}}
\left[a_n-a_n^\dagger\right]\cos\left(n\sigma\over
2\alpha\right),\cr
p(\sigma)&={1\over 2\pi\alpha}\left[\sqrt{{{\bw}_0\over
2\alpha^\prime}}(a_0+a_0^\dagger)+
\sum_{n=1}^\infty \sqrt{{\bw_n\over
\alpha^\prime}}
\left[a_n+a_n^\dagger\right]\cos\left(n\sigma\over
2\alpha\right)\right].}}

\noindent
2)Dirichlet Boundary Condition:
\eqn\openexpb{\eqalign{
x(\sigma)&=
\sum_{n=1}^\infty i\sqrt{{\alpha^\prime\over\bw_n}}
\left[a_{-n}-a_{-n}^\dagger\right]\sin\left(n\sigma\over
2\alpha\right),\cr
p(\sigma)&={1\over 2\pi\alpha}\sum_{n=1}^\infty\sqrt{{\bw_n\over
\alpha^\prime}}
\left[a_{-n}+a_{-n}^\dagger\right]\sin\left(n\sigma\over
2\alpha\right).}}

Given this mode expansion it is now straightforward to write the
continuity equations \conti\ in terms of modes. A dramatic
simplification occurs in the large $\mu$ limit, since then
$\w_{n(i)}=\bw_{n(i)}\simeq \mu\alpha_{(i)}$, where
we take without loss of generality $\alpha_{(1)}=y, \alpha_{(2)}=1-y$
and $\alpha_{(3)}=-1$(if there is only a single incoming string
$y=1$). The idea is to isolate the oscillators of string 3
and write them in terms of oscillators in string 1 and 2 by
multiplying \conti\ by a complete, orthonormal basis of functions
along string 3 and then integrate over the strip.
The equations reduce to
\eqn\contioscilla{\eqalign{
\left(A_{n(3)}-A^\dagger_{n(3)}-{C_{mn}^{1}\over
\sqrt{y}}(A_{n(1)}-A^\dagger_{n(1)})-{C_{mn}^{2}\over
\sqrt{1-y}}(A_{n(2)}-A^\dagger_{n(2)})\right)|V\ra&=0,\cr
\left(A_{n(3)}+A^\dagger_{n(3)}+{C_{mn}^{1}\over
\sqrt{y}}(A_{n(1)}-A^\dagger_{n(1)})+{C_{mn}^{2}\over
\sqrt{1-y}}(A_{n(2)}-A^\dagger_{n(2)})\right)|V\ra&=0.}}
$A_{n(r)}$ can be either $a_{n(r)},a_{-n(r)}$ or $\alpha_{n(r)}$
depending on whether the $r$-th string is an open string with a Neumann,
Dirichlet boundary condition or a closed string.
$C_{mn}^1(C_{mn}^2)$ is the integral over the domain of string
1(2) of the
product of Fourier modes of string 3 and string 1(2). One can easily
solve \contioscilla , which yield the following large $\mu$ Neumann 
matrices:
\eqn\Neummmm{
N^{(13)}_{m,n}=-{C_{mn}^1\over \sqrt{y}},\qquad
N^{(23)}_{m,n}=-{C_{mn}^2\over \sqrt{1-y}}.}
 From \conti\ one can also show that $N^{(33)}, N^{(11)},N^{(12)}$ and
$N^{(22)}$ are of order ${1\over \mu}$ and therefore suppressed in the
large $\mu$ limit.

We now make explicit the large $\mu$ Neumann matrices for the
interactions considered in this paper.

\appendix{C}{Large $\mu$ Neumann Matrices and Prefactor for
$H_{o\leftrightarrow oo}$}

The parametrization of the Mandelstam diagram in Fig. 1 is:
\eqn\intervalooo{\eqalign{
\sigma_3 &=\sigma, \qquad\qquad\quad 0\leq \sigma\leq 2\pi,\cr
\sigma_{1}&=\sigma, \qquad\qquad\quad 0\leq \sigma\leq 2\pi y,\cr
\sigma_{2}&=\sigma-2\pi y, \qquad 2\pi y\leq \sigma\leq 2\pi.}}
Therefore, the Neumann matrices are ($m,n>0$)\foot{If one of the
indices is zero,
we have to divide the
present formula by $\sqrt{2}$. Note that compared to \HeZU\ there
are some small sign differences in the Neumann matrices involving
string 3 due to our choice of parametrization of the worldsheet.}:

\noindent
1)Neumann Boundary Condition:
\eqn\fourierover{\eqalign{
{\bar N}^{(13)}_{m,n}(y)&\hskip-2pt\simeq\hskip-2pt-\hskip-1pt
{1\over \pi\sqrt{y}}\int_0^{2\pi y}d\sigma
\cos\left({n\sigma\over
2}\right)\cos\Bigl({m\sigma\over 2y}\Bigr)={2(-1)^{m+1}n\sin(n\pi 
y)\over
\pi\sqrt{y}(n^2\hskip-2pt-\hskip-2pt m^2/y^2)},\cr
{\bar N}^{(23)}_{m,n}(y)&\hskip-2pt\simeq\hskip-2pt-\hskip-1pt
{1\over \pi\sqrt{1
\hskip-2pt-\hskip-2pt y}}\int_{2\pi
y}^{2\pi } d\sigma\cos\left({n\sigma\over
2}\right)\cos\Bigl({m(\sigma\hskip-2pt-\hskip-2pt 2\pi y)\over
2(1\hskip-2pt-\hskip-2pt y)}\Bigr)\hskip-2pt
=\hskip-2pt{2(-1)^{n+1}n\sin(n\pi(1\hskip-2pt
-\hskip-2pt y))\over
\pi\sqrt{1\hskip-2pt-\hskip-2pt y}(n^2\hskip-2pt-\hskip-2pt
m^2/(1\hskip-2pt-\hskip-2pt y)^2)},}}

\noindent
2)Dirichlet Boundary Condition:
\eqn\fourieroverdi{\eqalign{
{\bar N}^{(13)}_{-m,-n}(y)&\hskip-2pt\simeq\hskip-2pt-\hskip-1pt
{1\over \pi\sqrt{y}}\hskip-2pt\int_0^{2\pi y}d\sigma
\sin\left({n\sigma\over
2}\right)\sin\Bigl({m\sigma\over 2y}\Bigr)\hskip-2pt
=\hskip-2pt{2(-1)^{m+1}m\sin(n\pi 
y)\over
\pi y^{3/2}(n^2-m^2/y^2)},\cr
{\bar N}^{(23)}_{-m,-n}(y)&\hskip-2pt\simeq\hskip-2pt-\hskip-1pt
{1\over \pi\sqrt{1\hskip-2pt-\hskip-2pt y}}\hskip-2pt\int_{2\pi
y}^{2\pi }\hskip-2pt d\sigma \sin\left({n\sigma\over
2}\right)\sin\Bigl({m(\sigma\hskip-2pt-\hskip-2pt 2\pi y)\over
2(1\hskip-2pt-\hskip-2pt y)}\Bigr)\hskip-2pt=\hskip-2pt
{2(-1)^{n+1}m\sin(n\pi (1\hskip-2pt-\hskip-2pt y))\over
\pi(1\hskip-2pt-\hskip-2pt y)^{3/2}
(n^2\hskip-2pt-\hskip-2pt m^2/(1\hskip-2pt-\hskip-2pt y)^2)}.}}

The prefactor can be obtained via factorization of the Neumann
matrices. The leading large $\mu$ prefactor is
\foot{Also we have to divide by $\sqrt{2}$ for $n=0$. Apart
from the sign difference in ${\bar F}_{n(3)}$ due to our worldsheet
parametrization, these quantities are those in \HeZU\ up to an overall
numerical factor.}
$(n>0)$
\eqn\Ftilplus{\eqalign{
&{\bar F}_{n(3)}(y)\simeq {-\sqrt{2}
n\sin(\pi ny)\over2\pi \mu^2}.}}

\appendix{D}{Large $\mu$ Neumann Matrices and Prefactor for
$H_{o\leftrightarrow c}$}

The parametrization of the Mandelstam diagram in Fig. 9 is:
\eqn\intervaloc{\eqalign{
\sigma_o &=\sigma, \qquad\quad\quad  0\leq \sigma\leq 2\pi,\cr
\sigma_{c}&=\sigma, \qquad\quad\quad 0\leq \sigma\leq 2\pi.}}
The Neumann matrices are ($m>0$)

\noindent
1)Neumann Boundary Condition:
\eqn\fourieroveroc{
{N}^{(oc)}_{m,n}\simeq-{\sqrt{2} \over 2\pi}\int_0^{2\pi}
d\sigma e^{-in\sigma}\cos\left({m\sigma\over
2}\right)=\Bigg\{\matrix{-{1\over
\sqrt{2}}(\delta_{m,2n}+\delta_{m,-2n})\ &m:\hbox{even}\cr
{8in\over \sqrt{2}\pi(4n^2-m^2)}\ &m:\hbox{odd}},}

\noindent
2)Dirichlet Boundary Condition:
\eqn\fourieroverdioc{
{N}^{(oc)}_{-m,n}\simeq-{\sqrt{2}\over 2\pi}\int_0^{2\pi }
d\sigma e^{-in\sigma}\sin\left({m\sigma\over 2}\right)
=\Bigg\{\matrix{{i\over
\sqrt{2}}(\delta_{m,2n}-\delta_{m,-2n})\ &m:\hbox{even}\cr
{-4m\over \sqrt{2}\pi(m^2-4n^2)}\ &m:\hbox{odd}}.}

The prefactor can be obtained via factorization of the Neumann
matrices. The leading large $\mu$ prefactor is
\eqn\Ftilplus{\eqalign{
&{F}_{n(c)}\simeq {n\over 2\pi i\mu^2}.}}

\appendix{E}{Large $\mu$ Neumann Matrices and Prefactor for
$H_{c\leftrightarrow c}$}

The closed string rearrangement interaction

\ifig\Seminearest{Closed string rearrangement interaction diagram.}
{\epsfxsize3in\epsfbox{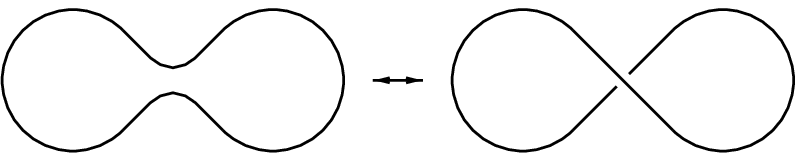}}

\noindent
has  the following worldsheet parametrization 
\ifig\Seminearest{Closed string rearrangement  parametrization.}
{\epsfxsize2in\epsfbox{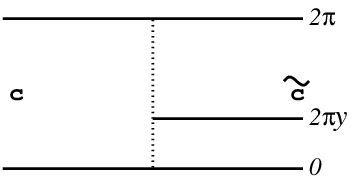}}

\eqn\interval{\eqalign{
\sigma_c &=-\sigma, \qquad\qquad\quad\quad\;\;\;
 0\leq \sigma\leq 2\pi,\cr
\sigma_{\tilde c}&=\left\{\matrix{\hskip-50pt\sigma
& 0\leq \sigma\leq 2\pi y\cr
2\pi(1+y)-\sigma&\; 2\pi y\leq \sigma\leq 2\pi}\right..}}
The Neumann matrices are:

\eqn\fourierover{\eqalign{
{N}^{({\tilde c}c)}_{m,n}(y)&=-{1\over 2\pi}\left[\int_{0}^{2\pi
y}d\sigma e^{i(n-m)\sigma}+e^{2\pi iny}\int_{0}^{2\pi
(1-y)}d\sigma e^{i(n+m)\sigma}\right]\cr&={i\over 2\pi}
\left[{e^{2\pi i(n-m)y}-1\over n-m}
+{e^{-2\pi imy}-e^{2\pi iny}\over n+m}\right].}}

The prefactor can be obtained via factorization of the Neumann
matrices and  can be written, just as in $H_{c\leftrightarrow cc}$.

\appendix{F}{Hamiltonian with Bosonic External States}
Since on the gauge theory side we only consider the two-point
function with bosonic impurities, on the string field theory side we
also restrict ourselves to the Hamiltonian matrix elements of purely
bosonic excitations on the vacuum state.
After this restriction we can simplify the light-cone Hamiltonian of
pp-wave open-closed string field theory quite a lot, as is the case
\LeeRM\ for the pp-wave closed string field theory \SpradlinAR.
In this appendix we shall reduce the light-cone Hamiltonian of the
cubic open string vertex on the pp-wave background \StefanskiZC\ to
our expression \cubicopen.

To diagonalize the free light-cone Hamiltonian, the dynamical
variables in this theory are rewritten by
\eqn\diag{\eqalign{&x_{n(u)}^I=i\sqrt{\al'\over 2\omega_{n(u)}}
(a_{n(u)}^I-a_{n(u)}^{I\dagger}),\qquad
p_{n(u)}^I=\sqrt{\omega_{n(u)}\over 2\al'}
(a_{n(u)}^I+a_{n(u)}^{I\dagger}),\cr
&R_{\pm 0(u)}^A={1\over 2}
(1\mp ie(\al_u)\Om\Pi)^A{}_B\lambda_{0(u)}^B,\qquad
R_{\pm 0(u)A}={1\over 2}(1\mp ie(\al_u)\Om\Pi)_A{}^B\theta_{0(u)B},}}
in terms of bosonic oscillators $a$ and fermionic ones $R$
\eqn\oscillators{\eqalign{
&[a_{n(u)}^I,a_{m(v)}^{J\dagger}]
=\delta^{IJ}\delta_{nm}\delta_{uv},\cr
&\{R_{\pm 0A(u)},R_{\pm 0(v)}^B\}=0,\qquad
\{R_{\pm 0A(u)},R_{\mp 0(v)}^B\}
=\delta_{uv}{1\over 2}(1\mp i\Om\Pi)_A{}^B.}}
We have omitted the redefinition of the fermionic non-zero modes
because it is not necessary for our analysis later.
Various gamma matrices here are given by
\eqn\Ome{\Om=\ga^{78}=\pmatrix{\Om_A{}^B&0\cr0&\Om^A{}_B}
=\pmatrix{i1_4&0\cr0&-i1_4},}
\eqn\Pai{\Pi=\ga^{1234}=\pmatrix{\Pi_A{}^B&0\cr0&\Pi^A{}_B}.}
Since $\tr\Pi=0$, $\Pi^2=1_8$ and $\Pi_A{}^B=\Pi^B{}_A$, hereafter we
can choose
\eqn\pai{\Pi_A{}^B=\Pi^B{}_A=\pmatrix{-1_2&0\cr0&1_2}.}

In terms of these oscillators, the interaction light-cone Hamiltonian
on the pp-wave background is\foot{As in \StefanskiZC\ the overall
normalization of the interaction Hamiltonian is not determined by
supersymmetry algebra. Although in the main text we fix the
normalization from the gauge theory result, in this appendix we simply
follow the normalization of \StefanskiZC.}
\eqn\Hamiltonian{|H\ra=(1-4\mu\alpha K)^{1/2}
\Bigl[\sqrt{\alpha'}{\cal K}^L
+{(\alpha')^{3/2}\over 2\sqrt{2}\alpha}{\cal K}^i\rho^{iCD}Y_CY_D
+{(\alpha')^{5/2}\over 24\alpha^2}
{\cal K}^R\epsilon^{CDEF}Y_CY_DY_EY_F\Bigr]|V\ra,}
with the sum of $i$ running over the $SO(6)$ indices.
The prefactor part is given with
\eqn\K{\eqalign{&{\cal K}^i=P^i-i\mu{\alpha\over\alpha'}R^i
+\sum_{u=1}^3\sum_{n=1}^\infty{\bar F}_{n(u)}
a_{n(u)}^{\dagger i},\cr
&{\cal K}^{L,R}=\sum_{u=1}^3\sum_{n=1}^\infty
{\bar F}_{-n(u)}a_{-n(u)}^{\dagger L,R},}}
and
\eqn\Y{\eqalign{Y_A&=(1-4\mu\alpha K)^{-1/2}
(1-2\mu\alpha K(1+i\Omega\Pi))_A{}^B{\cal Y}_B,\cr
{\cal Y}_A&=-{\alpha\over\alpha'}\Theta_A
+\sum_{r=1}^3\sum_{n=1}^\infty G_{n(r)A}{}^BR_{-n(r)B}.}}
And the overlapping part $|V\ra$ is given by
\eqn\overlap{|V\ra=|V_{\rm bos}\ra|V_{\rm ferm}\ra
|\al_3|\delta\biggl(\sum_{r=1}^3\alpha_r\biggr),}
with
\eqn\Vb{|V_{\rm bos}\ra=\exp\biggl\{{1\over 2}
\sum_{r,s=1}^3\sum_{m,n}a_{m(u)}^{I\dagger}
{\bar N}_{m,n\;IJ}^{(rs)}a_{n(v)}^{J\dagger}\biggr\},}
\eqn\Vf{|V_{\rm ferm}\ra=\exp\biggl\{
\sum_{r,s=1}^3\sum_{m,n=1}^\infty
R_{-m(r)}^AQ_{m,n\;A}^{(rs)}{}^BR_{-n(s)B}
-\sum_{r=1}^3\sum_{m=1}^\infty
R_{-m(r)}^AQ_{m\;A}^{(r)}{}^B\Theta_B\biggr\}|V_{\rm ferm}^0\ra,}
\eqn\Vfzero{|V_{\rm ferm}^0\ra
=\prod_{A=1}^4\biggl(\sum_{r=1}^3\al_r\theta_{0(r)A}\biggr)|0\ra_{123}.}
Here the state $|0\ra_r$ is the $SU(4)$ invariant state, which means
$(m>0)$
\eqn\ground{\eqalign{
a_{n(r)}^I|0\ra_r=0,\quad
& R_{m(r)}^A|0\ra_r=0,\quad R_{m(r)A}|0\ra_r=0,\cr
&\lambda_{0(r)}^A|0\ra_r=0.}}
We can relate it to the vacuum state $|\vac\ra$, $(m>0)$
\eqn\vacuum{\eqalign{
a_{n(r)}^I|\vac\ra_r=0,\quad
& R_{m(r)}^A|\vac\ra_r=0,\quad R_{m(r)A}|\vac\ra_r=0,\cr
& R_{+0(r)}^A|\vac\ra_r=0,\quad R_{+0(r)A}|\vac\ra_r=0,}}
by $(r=1,2)$
\eqn\relation{|0\ra_r=R_{-0(r)}^3R_{-0(r)}^4|\vac\ra_r,\qquad
|0\ra_3=R_{-0(3)}^1R_{-0(3)}^2|\vac\ra_3.}

Since we do not have fermionic excitations in the external states, the
squeezed state of fermionic non-zero modes does not contribute.
Furthermore, to cancel six fermionic zero mode creation operators, the
only possibility is to use four fermionic annihilation operators in
$|V_{\rm ferm}^0\ra$ and two in the prefactor.
After cancelling out the fermionic zero modes, we have
\eqn\Hamil{|H\ra=C_S{\cal K}^{z'}
\exp\biggl\{{1\over 2}\sum_{r,s=1}^3\sum_{m,n}
a_{m(u)}^{I\dagger}\bar N_{m,n\;IJ}^{(rs)}a_{n(v)}^{J\dagger}\biggr\}
|\vac\ra_{123},}
with
\eqn\C{C_S={-\al\al_3^2\over 2\sqrt{\al'}}{1\over\sqrt{1-4\mu\al K}}
\simeq\sqrt{\pi\mu}\bigl(y(1-y)\bigr)^{3/2},}
for the interaction part of the light-cone Hamiltonian.
Here we have used that
\eqn\Zprime{{\cal K}^i\rho^{i34}=\sqrt{2}{\cal K}^{z'}.}
As in \chargevac\susyal\ this fact follows directly from $U(1)_{56}$
symmetry and can also be explicitly confirmed from the following
representation of the gamma matrices with some additional conditions
like antisymmetric property $\rho^i_{AB}=-\rho^i_{BA}$ and \pai:
\eqn\rhosix{\eqalign{
\rho^1=\pmatrix{0&0&0&1\cr0&0&-1&0\cr0&1&0&0\cr-1&0&0&0},&\qquad
\rho^2=\pmatrix{0&0&0&-i\cr0&0&-i&0\cr0&i&0&0\cr i&0&0&0},\cr
\rho^3=\pmatrix{0&0&1&0\cr0&0&0&1\cr-1&0&0&0\cr0&-1&0&0},&\qquad
\rho^4=\pmatrix{0&0&i&0\cr0&0&0&-i\cr-i&0&0&0\cr0&i&0&0},\cr
\rho^5=\pmatrix{0&-1&0&0\cr1&0&0&0\cr0&0&0&1\cr0&0&-1&0},&\qquad
\rho^6=\pmatrix{0&-i&0&0\cr i&0&0&0\cr0&0&0&-i\cr0&0&i&0}.}}

\appendix{G}{Neumann Coefficients for the Open-Closed Transition
Vertex}
The purpose of this appendix is to derive the bosonic Neumann
coefficients for the open-closed transition vertex and prove a
decomposition theorem for it \SchwarzBC\PankiewiczGS: ($r,s=o,c$)
\eqn\factorize{N^{(rs)}_{m,n}=
{ZY_{m(r)}Y_{n(s)}\over(\Om_r)_m/\al_r+(\Om_s)_n/\al_s}.}
In this appendix, we shall set $\al_o=-\al_c=1$.
Since the Neumann coefficients only depend on the ratio of momentum,
our simplification does not reduce any information.
The reason for proving it is as follows.
In the functional interpretation, the bosonic vector constituent of
the prefactor is given as \GreenFU
\eqn\pref{\sum_{r}\sum_{m}F_{m(r)}\al_{m(r)}^\dagger|V\ra
\sim\lim_{\sigma\to 0}\sqrt{\sigma}
\Bigl(2\pi P^o(\sigma)\pm\partial X^o(\sigma)\Bigr)|V\ra.}
Therefore, if we can prove the decomposition theorem \factorize, it
will suggest that
\eqn\FeqY{F_{m(r)}\sim Y_{m(r)}.}

For readers who are not interested in the details of the construction
and calculation, we summarize the main result here.
In the large $\mu$ limit, the behavior of various Neumann coefficients
with Neumann boundary condition for the open string is given by
\eqn\ocsummary{\eqalign{
&F^{c}_{n}\sim{n\over 2\pi i\mu^2},\quad
N^{(oo)}_{m,n}\simeq 0,\quad N^{(cc)}_{m,n}\simeq 0,\cr
&N^{(oc)}_{m,n}\simeq
-{1\over 2\pi}\int_0^{2\pi}d\sigma\sqrt{2}\cos{m\sigma\over 2}
e^{-in\sigma}=\left\{\matrix{-(1/\sqrt{2})\delta_{m,2n}
&\hbox{for}\;m:\hbox{even}\cr
4i\sqrt{2}n/[\pi(4n^2-m^2)]&\!\hbox{for}\;m:\hbox{odd}}\right..}}
Here we have rewritten the result of the following derivation in
$\sin/\cos$ basis into the $\exp$ basis, since it is more convenient
in the comparison with the gauge theory result.
Also, $\simeq$ denotes the large $\mu$ behavior (as we have used in
the whole paper), while $\sim$ means uncertainty of the overall
normalization.

Similarly, the large $\mu$ Neumann coefficients with Dirichlet
boundary condition for the open string is given by
\eqn\diri{\eqalign{
&N^{(oo)}_{m,n}\simeq 0,\quad N^{(cc)}_{m,n}\simeq 0,\cr
&N^{(oc)}_{m,n}\simeq
-{1\over 2\pi}\int_0^{2\pi}d\sigma\sqrt{2}\sin{m\sigma\over 2}
e^{-in\sigma}=\left\{\matrix{(i/\sqrt{2})\delta_{m,2n}
&\hbox{for}\;m:\hbox{even}\cr
-2\sqrt{2}m/[\pi(m^2-4n^2)]&\!\hbox{for}\;m:\hbox{odd}}\right..}}

Let us begin with the construction of the Neumann coefficients.
Neumann coefficients appear as the oscillator representation of the
overlapping condition in the interaction vertex.
Therefore, in the momentum space the squeezed state is simply the
(infinite-dimensional) local momentum conservation condition on the
worldsheet.
\eqn\deltamom{\Delta\Bigl(P^c(\si)+P^o(\si)\Bigr).}
In mode expansion this is written more explicitly as
\eqn\deltaosc{\Delta\Bigl(p_{(c)}+U_{co}p_{(o)}\Bigr).}
Note here that we omit the mode indices.
More precisely, $p_{(c)}$ and $p_{(o)}$ denote infinite dimensional
column vector and $U_{co}$ is the infinite dimensional transformation
matrix.
Since $p_{(c)}$ and $p_{(o)}$ contain the same information of the same
string except at the open string boundary, $U_{co}$ should be an
orthonormal matrix, which means\foot{In the case of cubic closed
string vertex, this orthonormality also holds up to some
normalization.
In this case third string momentum relates to first and second
string ones by
\eqn\three{p_{(3)}=\Bigl(X^{(31)}\quad X^{(32)}\Bigr)
\pmatrix{p_{(1)}\cr p_{(2)}},}
and if we define
\eqn\U{U=\Bigl(\sqrt{-\al_1/\al_3}X^{(31)}
\quad\sqrt{-\al_2/\al_3}X^{(32)}\Bigr),}
then $U$ enjoys the orthonormality.}
\eqn\ortho{U_{oc}U_{co}=1_{oo},\qquad U_{co}U_{oc}=1_{cc},}
if we define $U_{oc}=U_{co}^\T$.

To state the interaction vertices in terms of eigenstates of the free
Hamiltonian, we have to express them in the oscillator space.
For this purpose, we need to transform the momentum eigenstates into
number eigenstates by
\eqn\pton{|p\ra=\sum_n{(a^\dagger)^n\over\sqrt{n!}}|0\ra
\la 0|{a^n\over\sqrt{n!}}|p\ra=({\rm const})\times\psi(p)|0\ra}
with
\eqn\wave{\psi(p)=\exp\biggl(-{1\over 2}p^\T{1\over\Om}\;p
+\sqrt{2}a^{\dagger\T}{1\over\sqrt{\Om}}\;p
-{1\over 2}a^{\dagger\T}a^\dagger\biggr)}
where $\Om$ is the diagonal frequency $\om$ matrix.
Using this wave function, the overlapping condition is then given as
\eqn\exponent{\int dp_{(c)}dp_{(o)}\psi_c(p_{(c)})\psi_o(p_{(o)})
\Delta(p_{(c)}+U_{co}p_{(o)})|0\ra_{oc}.}
After performing the integration, we find
\eqn\V{\exp\biggl[{1\over 2}
\Bigl(a_{(c)}^{\dagger}\;a_{(o)}^{\dagger}\Bigr)
N\biggl({a_{(c)}^{\dagger}\atop a_{(o)}^{\dagger}}\biggr)\biggr]
|0\ra_{oc},}
with the Neumann coefficients given by
\eqn\N{\eqalign{
N^{(oo)}&=-1+2{1\over\sqrt{\Om_o}}
{1\over U_{oc}\Om_c^{-1}U_{co}+\Om_o^{-1}}{1\over\sqrt{\Om_o}},\cr
N^{(oc)}&=-2{1\over\sqrt{\Om_o}}
{1\over U_{oc}\Om_c^{-1}U_{co}+\Om_o^{-1}}U_{oc}
{1\over\sqrt{\Om_c}},\cr
N^{(cc)}&=-1+2{1\over\sqrt{\Om_c}}U_{co}
{1\over U_{oc}\Om_c^{-1}U_{co}+\Om_o^{-1}}U_{oc}{1\over\sqrt{\Om_c}}.}}
Hence, if we define $\Ga$ as
\eqn\DGam{\Ga=\Om_o+U_{oc}\Om_cU_{co},}
the Neumann coefficients are given by
\eqn\DN{\eqalign{N^{(oo)}&=2\sqrt{\Om_o}
\biggl({1\over 2\Om_o}-{1\over\Ga}\biggr)\sqrt{\Om_o},\cr
N^{(oc)}&=-2\sqrt{\Om_o}
\biggl({1\over\Ga}U_{oc}\biggr)\sqrt{\Om_c},\cr
N^{(cc)}&=2\sqrt{\Om_c}
\biggl({1\over 2\Om_c}-U_{co}{1\over\Ga}U_{oc}\biggr)\sqrt{\Om_c},}}
where we have used the relation
\eqn\part{
{1\over 1+\sqrt{\Om_o}U_{oc}\Om_c^{-1}U_{co}\sqrt{\Om_o}}
=1-\sqrt{\Om_o}{1\over\Om_o+U_{oc}\Om_cU_{co}}\sqrt{\Om_o},}
because of the following identity:
\eqn\inverse{{1\over 1+X^\T X}=1-X^\T{1\over 1+XX^\T}X.}
Since we have
\eqn\asympOm{\Om_c\simeq\mu 1_{cc},\qquad\Om_o\simeq\mu 1_{oo},}
in the large $\mu$ limit, we have
\eqn\asympinv{\Ga\simeq 2\mu 1_{oo},}
and therefore, together with \ortho\ we can derive the asymptotical
behavior of Neumann coefficients without any difficulty:
\eqn\asympN{N^{(oo)}\simeq 0,\qquad
N^{(oc)}\simeq -U_{oc},\qquad
N^{(cc)}\simeq 0.}
Note that the arguments so far are quite general.
Even if we consider other interaction vertices, the fact that the
Neumann coefficients reduce to the orthonormal overlapping matrix is
still the case, because the orthonormality \ortho\ does not depend on
the vertices we are discussing.

Let us concentrate on the open-closed transition vertex from now on.
Compared with the current discussion for \asympN, our understanding on
the decomposition theorem is slightly insufficient.
Although the techniques we shall apply are more or less similar for
all the vertices, we have to discuss each case separately.

\noindent$\bullet$
{\it Neumann Boundary Condition for the Open String}

We shall begin with the open-closed transition vertex with the
Neumann boundary condition for the open string.
Our plan is first to work out the oscillator expansion $U_{co}$ of
closed string oscillators in terms of open string ones, and then prove
a decomposition theorem for the Neumann coefficients.

The local momentum conservation on the worldsheet in terms of the
oscillator expansion of each momentum is $(m>0)$
\eqn\modes{\eqalign{p_{0(c)}&=-p_{0(o)},\cr
p_{m(c)}&=-p_{2m(o)},\cr
p_{-m(c)}&=\sum_{n=1}^\infty{-8m\over\pi(4m^2-(2n-1)^2)}p_{2n-1(o)}.}}
Here we have used the following formula:
\eqn\cossin{\eqalign{&{1\over 2\pi\al}\int_0^{2\pi\al}d\sigma\;
2\cos{m\si\over\al}\cos{n\si\over\al}
=\delta_{m,n}+\delta_{m,-n},\cr
&{1\over 2\pi\al}\int_0^{2\pi\al}d\sigma\;
2\sin{m\si\over\al}\sin{n\si\over\al}
=\delta_{m,n}+\delta_{m,-n},\cr
&{1\over 2\pi\al}\int_0^{2\pi\al}d\sigma\;
2\cos{m\si\over\al}\cos{n\si\over 2\al}
=\delta_{2m,n}+\delta_{2m,-n},\cr
&{1\over 2\pi\al}\int_0^{2\pi\al}d\sigma\;
2\sin{m\si\over\al}\cos{n\si\over 2\al}
=\Biggl\{\matrix{0&\hbox{for}\;n:\hbox{even}\cr
{\displaystyle 8m\over\displaystyle\pi(4m^2-n^2)}
&\!\hbox{for}\;n:\hbox{odd}}.}}
Hence $U_{co}$ is given as
\eqn\Uco{(U_{co})_{m,n}=\pmatrix{\delta_{2m,n}&0\cr
0&-8m/\bigl[\pi(4m^2-n^2)\bigr]},}
where the left (right) column denotes the non-negative (negative)
modes of the closed string, while the upper (lower) row denotes the
even (odd) modes.
The upper-left block is just trivial Kronecker's delta overlap.

Next, let us proceed to derive the decomposition theorem for the
Neumann coefficients.
As preliminaries, we can derive the following identities:
\eqn\coc{U_{co}(\Om_o^2-\mu^2)U_{oc}=\Om_c^2-\mu^2,}
\eqn\cocinv{U_{co}{1\over\Om_o^2-\mu^2}U_{oc}
={1\over\Om_c^2-\mu^2}+V_cV_c^\T,}
\eqn\ocoinv{U_{oc}{1\over\Om_c^2-\mu^2}U_{co}
={1\over\Om_o^2-\mu^2}-V_oV_o^\T,}
with $V_o$ and $V_c$ defined by
\eqn\Vo{\bigl(V_o\bigr)_n=\left\{\matrix{0&n:\hbox{even}\cr
4\sqrt{2}/(\pi n^2)&\!\!n:\hbox{odd}}\right.,}
\eqn\Vc{\bigl(V_c\bigr)_n=\left\{\matrix{0&n\ge 0\cr
-\sqrt{2}/n&n<0}\right..}
These two vectors are related by
\eqn\VcVo{V_c=U_{co}V_o,\qquad V_o=U_{oc}V_c.}
Note that in \cocinv\ and \ocoinv\ both the LHS and RHS are
divergent for the zero mode.
However, since it decouples completely from other modes, we can
define the zero mode for $V_o$ and $V_c$ in \Vo\ and \Vc\ as other
trivial modes formally.
These two identities \cocinv\ and \ocoinv\ are very useful in our
following analysis.

The standard strategy to derive the decomposition theorem is to
rewrite $\Ga$ as
\eqn\Gam{\Gamma=\Om_o+U_{oc}\Om_cU_{co}
=\Om_o+\mu+U_{oc}(\Om_c-\mu)U_{co}=\Om_o-\mu+U_{oc}(\Om_c+\mu)U_{co},}
and compute
\eqn\multiply{\eqalign{&\bigl(\Gamma-(\Om_o+\mu)\bigr)
{1\over\Om_o^2-\mu^2}\bigl(\Gamma-(\Om_o-\mu)\bigr)\cr
&=U_{oc}(\Om_c-\mu)U_{co}{1\over\Om_o^2-\mu^2}
U_{oc}(\Om_c+\mu)U_{co}\cr
&=1+U_{oc}(\Om_c-\mu)V_cV_c^\T(\Om_c+\mu)U_{co},}}
with the help of \cocinv.
Expanding the LHS of \multiply\ and multiplying by $\Ga^{-1}$ both on
the left and right, we find
\eqn\matoo{{1\over\Om_o^2-\mu^2}
={1\over\Ga}{1\over\Om_o-\mu}
+{1\over\Om_o+\mu}{1\over\Ga}
+{1\over\Ga}U_{oc}(\Om_c-\mu)V_c
V_c^\T(\Om_c+\mu)U_{co}{1\over\Ga}.}
Furthermore, after multiplying $(\Om_o+\mu)$ and $(\Om_o-\mu)$ on the
left and right respectively and solving for $\Ga^{-1}$, we find the
decomposition theorem for $N^{(oo)}$:
\eqn\decoo{\biggl({1\over 2\Om_o}-{1\over\Ga}\biggr)_{mn}
={\bigl[(\Om_o+\mu)\Ga^{-1}U_{oc}(\Om_c-\mu)V_c\bigr]_m
\bigl[(\Om_o-\mu)\Ga^{-1}U_{oc}(\Om_c+\mu)V_c\bigr]_n
\over(\Om_o)_m+(\Om_o)_n}.}
For other components, we need another multiplication:
\eqn\mult{\bigl(\Ga-(\Om_o+\mu)\bigr){1\over\Om_o^2-\mu^2}U_{oc}
=U_{oc}(\Om_c-\mu)U_{co}{1\over\Om_o^2-\mu^2}U_{oc}
=U_{oc}{1\over\Om_c+\mu}
+U_{oc}(\Om_c-\mu)V_cV_c^\T.}
Hence, we have
\eqn\matoc{{1\over\Om_o^2-\mu^2}U_{oc}
={1\over\Ga}{1\over\Om_o-\mu}U_{oc}
+{1\over\Ga}U_{oc}{1\over\Om_c+\mu}
+{1\over\Ga}U_{oc}(\Om_c-\mu)V_cV_c^\T.}
Multiplying \matoo\ on the right with $U_{oc}$ and subtracting the
result with \matoc, the decomposition theorem for $N^{(oc)}$ reads
\eqn\decco{\biggl({1\over\Ga}U_{oc}\biggr)_{mn}
={\bigl[(\Om_o+\mu)\Ga^{-1}U_{oc}(\Om_c-\mu)V_c\bigr]_m
\bigl[-(\Om_c+\mu)U_{co}\Ga^{-1}(\Om_o-\mu)V_o\bigr]_n
\over(\Om_o)_m-(\Om_c)_n}.}
Here we have used the following rewriting.
\eqn\rewrite{\biggl(1-U_{co}{1\over\Ga}U_{oc}(\Om_c+\mu)\biggr)V_c
=U_{co}{1\over\Ga}\bigl(\Ga-U_{oc}(\Om_c+\mu)U_{co}\bigr)U_{oc}V_c
=U_{co}{1\over\Ga}(\Om_o-\mu)V_o.}
To find the decomposition theorem for $N^{(cc)}$, we need first to
exchange the sign of $\mu$ in \matoc\ and take the transpose of it,
\eqn\matco{U_{co}{1\over\Om_o^2-\mu^2}
=U_{co}{1\over\Om_o+\mu}{1\over\Ga}
+{1\over\Om_c-\mu}U_{co}{1\over\Ga}
+V_cV_c^\T(\Om_c+\mu)U_{co}{1\over\Ga},}
and then sum over the following three equations,
\matoo\ after multiplying with $U_{co}$ and $U_{oc}$ on the left and
right respectively, \matoc\ after multiplying with $-U_{co}$ on the
left and \matco\ after multiplying with $-U_{oc}$ on the right:
\eqn\matcc{{1\over\Om_c^2-\mu^2}
=U_{co}{1\over\Ga}U_{oc}{1\over\Om_c+\mu}
+{1\over\Om_c-\mu}U_{co}{1\over\Ga}U_{oc}
-U_{co}{1\over\Ga}(\Om_o+\mu)V_o
V_o^\T(\Om_o-\mu){1\over\Ga}U_{oc}.}
Hence, the decomposition theorem for $N^{(cc)}$ reads
\eqn\deccc{
\biggl({1\over 2\Om_c}-U_{co}{1\over\Ga}U_{oc}\biggr)_{mn}
={\bigl[-(\Om_c-\mu)U_{co}\Ga^{-1}(\Om_o+\mu)V_o\bigr]_m
\bigl[-(\Om_c+\mu)U_{co}\Ga^{-1}(\Om_o-\mu)V_o\bigr]_n
\over-(\Om_c)_m-(\Om_c)_n}.}

Here we have related the Neumann coefficient matrices to vector
quantities.
However, there remains some important issues.
Apparently the LHS of \decoo\ is symmetric,
while the symmetricity of the RHS is not obvious.
The necessary and sufficient condition for a matrix of the form
$V_1V_2^\T$ to be symmetric is that two vectors $V_1$ and $V_2$ are
parallel.
This is also important to prove our decomposition theorem.
Our next task is to confirm it.
Rewriting \matoo\ as
\eqn\equa{\eqalign{
&{1\over\Ga}U_{oc}(\Om_c-\mu)U_{co}=1-{1\over\Ga}(\Om_o+\mu)\cr
&\quad=(\Om_o-\mu){1\over\Ga}+
(\Om_o+\mu){1\over\Ga}U_{oc}(\Om_c-\mu)V_c
V_c^\T(\Om_c+\mu)U_{co}{1\over\Ga}(\Om_o-\mu),}}
and multiplying with $V_o$ on the right, we obtain\foot{To make the
arguments more parallel with the Dirichlet case, we need to multiply
with $U_{oc}(\Om_c+\mu)V_c$. However, this will give an indefinite
form $\Ga^{-1}U_{oc}(\Om_c^2-\mu^2)V_c$ on the LHS.}
\eqn\solve{{1\over\Ga}U_{oc}(\Om_c-\mu)V_c
={\Om_o-\mu\over 1-X(\Om_o+\mu)}W,}
with $W$ and $X$ defined by
\eqn\W{W={1\over\Ga}V_o,}
\eqn\X{X=V_c^\T(\Om_c+\mu)U_{co}{1\over\Ga}(\Om_o-\mu)V_o.}
Therefore, the two vectors on the RHS are expressed by
\eqn\twovecc{\eqalign{
(\Om_o+\mu){1\over\Ga}U_{oc}(\Om_c-\mu)V_c
&={\Om_o^2-\mu^2\over 1-X(\Om_o+\mu)}W,\cr
(\Om_o-\mu){1\over\Ga}U_{oc}(\Om_c+\mu)V_c
&={\bigl(1-2\mu X\bigr)
(\Om_o^2-\mu^2)\over 1-X(\Om_o+\mu)}W.}}
In this way the RHS of \matoo\ is shown to be symmetric.

The same story holds for the RHS of \deccc.
Rewriting \matcc, we find
\eqn\twoveco{\eqalign{
-(\Om_c-\mu)U_{co}{1\over\Ga}(\Om_o+\mu)V_o
&={-(\Om_c^2-\mu^2)\over 1+X(\Om_c-\mu)}U_{co}W,\cr
-(\Om_c+\mu)U_{co}{1\over\Ga}(\Om_o-\mu)V_o
&={-(1-2\mu X)(\Om_c^2-\mu^2)\over
1+X(\Om_c-\mu)}U_{co}W.}}
Consequently, the Neumann matrices are given as
\eqn\decompose{\eqalign{
N^{(oo)}_{m,n}
&={Z(Y_o)_m(Y_o)_n\over(\Om_o)_m/\al_o+(\Om_o)_n/\al_o},\cr
N^{(oc)}_{m,n}
&={Z(Y_o)_m(Y_c)_n\over(\Om_o)_m/\al_o+(\Om_c)_n/\al_c},\cr
N^{(cc)}_{m,n}
&={Z(Y_c)_m(Y_c)_n\over(\Om_c)_m/\al_c+(\Om_c)_n/\al_c},}}
with
\eqn\Y{\eqalign{
Y_o&={\sqrt{\Om_o}(\Om_o^2-\mu^2)\over
1-X(\Om_o+\mu)}W,\cr
Y_c&={-\sqrt{\Om_c}(\Om_c^2-\mu^2)\over
1+X(\Om_c-\mu)}U_{co}W,}}
and
\eqn\Z{Z=2(1-2\mu X).}
In this way, we have proved the decomposition theorem \decomp\ as we
promised.

To compare our results in this appendix with the gauge theory side, we
need the large $\mu$ behavior of the Neumann matrices.
First of all, we need to know how $X$ behaves as $\mu\to\infty$.
We can show that
\eqn\upper{\lim_{\mu\to\infty}{X\over\mu}=0.}
So let us assume\foot{Here we do not claim that $c$ should be finite.}
\eqn\lower{\lim_{\mu\to\infty}\mu X=c.}
Under this assumption, the non-trivial part of various Neumann
coefficients read
\eqn\asymptotic{\eqalign{
Y_o&\simeq{1\over\pi\sqrt{2\mu}(1-2c)},\cr
Y_c&\simeq{n\over\sqrt{2\mu}},\cr
Z&\simeq 2(1-2c),}}
in terms of the unknown $c$.
Hence the large $\mu$ behavior of other Neumann coefficients is
\eqn\selfinteract{\eqalign{
N^{(oo)}_{m,n}&\simeq{1\over 2\pi^2\mu^2(1-2c)},\cr
N^{(cc)}_{m,n}&\simeq-{(1-2c)mn\over 2\mu^2}.}}

\noindent$\bullet$
{\it Dirichlet Boundary Condition for the Open String}

Although the result in the rest of this appendix is not necessary
directly for our purpose because of the selection rule \Feynmmoc, let
us proceed to the case of Dirichlet boundary condition for
completeness.
Since we have
\eqn\Dcossin{\eqalign{
&{1\over 2\pi\al}\int_0^{2\pi\al}d\sigma\;
\sqrt{2}\sin{n\si\over 2\al}
=\Biggl\{\matrix{0&\hbox{for}\;n:\hbox{even}\cr
{\displaystyle 2\sqrt{2}\over\displaystyle\pi n}
&\!\hbox{for}\;n:\hbox{odd}},\cr
&{1\over 2\pi\al}\int_0^{2\pi\al}d\sigma\;
2\cos{m\si\over\al}\sin{n\si\over 2\al}
=\Biggl\{\matrix{0&\hbox{for}\;n:\hbox{even}\cr
{\displaystyle 4n\over\displaystyle\pi(n^2-4m^2)}
&\!\hbox{for}\;n:\hbox{odd}},\cr
&{1\over 2\pi\al}\int_0^{2\pi\al}d\sigma\;
2\sin{m\si\over\al}\sin{n\si\over 2\al}
=\delta_{2m,n}-\delta_{2m,-n},}}
the orthonormal matrix $U_{co}$ is defined as
\eqn\DU{\bigl(U_{co}\bigr)_{m,n}
=\pmatrix{\delta_{-2m,n}&0\cr
0&2\sqrt{2}/[\pi n]\cr
0&4n/\bigl[\pi(n^2-4m^2)\bigr]}.}
Here the first row of the closed string indices denotes the negative
modes, the second one denotes the zero mode and the third one denotes
the positive modes. Similarly the first column of the open string
indices denotes the even modes while the second one denote
the odd modes.
Since the first block of the overlapping matrix $U_{co}$ is trivial
and decouple from other modes, hereafter we shall explicitly drop this
part.
Before proceeding to the calculation, let us present here a useful
formula corresponding to \cocinv\ in the case of the Neumann boundary
condition:
\eqn\useful{U_{co}{1\over\Om_o^2-\mu^2}U_{oc}
=\pmatrix{\pi^2/3&F^\T\cr F&(\Om_c^2-\mu^2)^{-1}},}
with
\eqn\F{(F)_m=-{\sqrt{2}\over m^2}.}

The analysis for the Dirichlet case is almost parallel to the Neumann
one.
The main differences are that for the present case the zero mode of
the closed string also couples non-trivially and that the key identity
\solve\ to prove the parallelness is easier. (See the footnote there.)
As our standard method adopted in the previous case of Neumann
boundary condition, let us calculate
\eqn\Dmultiply{\eqalign{
&\bigl(\Ga-(\Om_o+\mu)\bigr){1\over\Om_o^2-\mu^2}
\bigl(\Ga-(\Om_o-\mu)\bigr)
=U_{oc}(\Om_c-\mu)U_{co}{1\over\Om_o^2-\mu^2}U_{oc}(\Om_c+\mu)U_{co}\cr
&\qquad=U_{oc}\pmatrix{0&0\cr 2\mu(\Om_c-\mu)F&1}U_{co}
=1-U_{oc}WV^\T U_{co},}}
Here $V$ and $W$ are defined as
\eqn\DVW{V=\pmatrix{1\cr 0},\qquad W=\pmatrix{1\cr -2\mu(\Om_c-\mu)F}.}
Therefore, we have
\eqn\Dmatoo{{1\over\Om_o^2-\mu^2}
-{1\over\Ga}{1\over\Om_o-\mu}-{1\over\Om_o+\mu}{1\over\Ga}
=-{1\over\Ga}U_{oc}WV^\T U_{co}{1\over\Ga}.}
Similarly, for other components we need some more summations:
\eqn\Dmultone{\bigl(\Ga-(\Om_o+\mu)\bigr){1\over\Om_o^2-\mu^2}
U_{oc}(\Om_c+\mu)
=U_{oc}(\Om_c-\mu)U_{co}{1\over\Om_o^2-\mu^2}U_{oc}(\Om_c+\mu)
=U_{oc}\bigl(1-WV^\T\bigr),}
\eqn\Dmulttwo{(\Om_c-\mu)U_{co}
{1\over\Om_o^2-\mu^2}\bigl(\Ga-(\Om_o-\mu)\bigr)
=(\Om_c-\mu)U_{co}{1\over\Om_o^2-\mu^2}U_{oc}(\Om_c+\mu)U_{co}
=\bigl(1-WV^\T\bigr)U_{co}.}
Hence, we have
\eqn\Dmatoc{{1\over\Om_o^2-\mu^2}U_{oc}(\Om_c+\mu)
-{1\over\Ga}{1\over\Om_o-\mu}U_{oc}(\Om_c+\mu)
={1\over\Ga}U_{oc}\bigl(1-WV^\T\bigr),}
\eqn\Dmatco{(\Om_c-\mu)U_{co}{1\over\Om_o^2-\mu^2}
-(\Om_c-\mu)U_{co}{1\over\Om_o+\mu}{1\over\Ga}
=\bigl(1-WV^\T\bigr)U_{co}{1\over\Ga}.}
Multiplying \Dmatoo\ with $(\Om_o+\mu)$ and $(\Om_o-\mu)$ on the left
and right respectively, we obtain
\eqn\Doo{\biggl({1\over 2\Om_o}-{1\over\Ga}\biggr)_{mn}
=-{\bigl[(\Om_o+\mu)\Ga^{-1}U_{oc}W\bigr]_m
\bigl[(\Om_o-\mu)\Ga^{-1}U_{oc}V\bigr]_n
\over(\Om_o)_m+(\Om_o)_n}.}
Similarly, using also \Dmatoc\ and \Dmatco, we find
\eqn\Doc{\biggl({1\over\Ga}U_{oc}\biggr)_{mn}
=-{\bigl[(\Om_o+\mu)\Ga^{-1}U_{oc}W\bigr]_m
\bigl[\bigl((\Om_c+\mu)U_{co}\Ga^{-1}U_{oc}-1\bigr)V\bigr]_n
\over(\Om_o)_m-(\Om_c)_n},}
\eqn\Dcc{\biggl({1\over 2\Om_c}-U_{co}{1\over\Ga}U_{oc}\biggr)_{mn}
=-{\bigl[\bigl((\Om_c-\mu)U_{co}\Ga^{-1}U_{oc}-1\bigr)W\bigr]_m
\bigl[\bigl((\Om_c+\mu)U_{co}\Ga^{-1}U_{oc}-1\bigr)V\bigr]_n
\over-(\Om_c)_m-(\Om_c)_n}.}
To see the symmetry of the matrices, we have to relate $W$ to $V$.
For this purpose, let us rewrite \Dmatoo\ into
\eqn\Dsym{\eqalign{
&-(\Om_o-\mu){1\over\Ga}+{1\over\Ga}U_{oc}(\Om_c-\mu)U_{co}
=1-(\Om_o+\mu){1\over\Ga}-{1\over\Ga}(\Om_o-\mu)\cr
&\qquad=-(\Om_o+\mu){1\over\Ga}
U_{oc}WV^\T U_{co}{1\over\Ga}(\Om_o-\mu),}}
and multiply with $U_{oc}V$ on the right.
As a consequence we obtain
\eqn\WV{(\Om_o+\mu){1\over\Ga}U_{oc}W
={1\over X}(\Om_o-\mu){1\over\Ga}U_{oc}V,}
with
\eqn\X{X=V^\T U_{co}{1\over\Ga}(\Om_o-\mu)U_{oc}V.}
With the help of the following identities
\eqn\rewrite{\eqalign{
\biggl((\Om_c+\mu)U_{co}{1\over\Ga}U_{oc}-1\biggr)V
&=-U_{co}(\Om_o-\mu){1\over\Ga}U_{oc}V,\cr
\biggl((\Om_c-\mu)U_{co}{1\over\Ga}U_{oc}-1\biggr)W
&=-U_{co}(\Om_o+\mu){1\over\Ga}U_{oc}W
=-{1\over X}U_{co}(\Om_o-\mu){1\over\Ga}U_{oc}V,}}
we can write down the decomposition theorem as follows:
\eqn\DNdecomp{\eqalign{
N^{(oo)}_{m,n}
&={Z(Y_o)_m(Y_o)_n\over(\Om_o)_m/\al_o+(\Om_o)_n/\al_o},\cr
N^{(oc)}_{m,n}
&={Z(Y_o)_m(Y_c)_n\over(\Om_o)_m/\al_o+(\Om_c)_n/\al_c},\cr
N^{(cc)}_{m,n}
&={Z(Y_c)_m(Y_c)_n\over(\Om_c)_m/\al_c+(\Om_c)_n/\al_c},}}
with
\eqn\DY{\eqalign{Y_o&=\sqrt{\Om_o}(\Om_o-\mu){1\over\Ga}U_{oc}V,\cr
Y_c&=-\sqrt{\Om_c}U_{co}(\Om_o-\mu){1\over\Ga}U_{oc}V,\cr
Z&=-{1\over X}.}}

\listrefs

\end